\documentclass[11pt]{revtex4}

\usepackage{graphicx}
\usepackage{amssymb}
\def\deltacp{\delta_{CP}}
\begin{document}
\title{DUSEL Theory White Paper}

\author{S.~Raby$^{1}$, T.~Walker$^{1,2,3}$,
K.S.~Babu$^4$, H.~Baer$^5$,  A.B.~Balantekin$^6$, V.~Barger$^6$, Z.~Berezhiani$^7$,
A.~de Gouvea$^8$, \\  R.~Dermisek$^9$,   A.~Dolgov$^{10}$, P.~Fileviez
Perez$^6$,  G.~Gabadadze$^{11}$, A.~Gal$^{12}$,  P.~Gondolo$^{13}$, \\
W.~Haxton$^{14}$, Y.~Kamyshkov$^{15}$, B.~Kayser$^{16}$,
E.~Kearns$^{17}$,  B.~Kopeliovich$^{18}$,  K.~Lande$^{19}$, \\
D.~Marfatia$^{20}$, R.N.~Mohapatra$^{21}$, P.~Nath$^{22}$,
Y.~Nomura$^{23}$, K.A.~Olive$^{24}$,  J.~Pati$^{25}$,
S.~Profumo$^{26}$, \\  R.~Shrock$^{27}$,
Z.~Tavartkiladze$^{28}$,
K.~Whisnant$^{29}$, L.~Wolfenstein$^{30}$ }

\affiliation{%
\small{
$^{1}$Department of Physics, The Ohio State
University, $^{2}$Department of Astronomy, The Ohio State University,
$^{3}$Center for Cosmology and AstroParticle Physics, The Ohio
State University, $^4$Department of Physics, Oklahoma State University, $^5$Department
of Physics, Florida State University, $^6$Department of Physics, The
University of Wisconsin Madison, $^{7}$Gran Sasso Laboratory, Italy,
$^8$Department of Physics and Astronomy, Northwestern University,
$^9$Institute for Advanced Study, Princeton University, $^{10}$ITEP,
Moscow and Ferrara University, Italy, $^{11}$New York University, NY,
$^{12}$Hebrew University, Israel, $^{13}$ Department of Physics,
University of Utah, $^{14}$Department of Physics, University of
Washington, Seattle, $^{18}$Department of Physics and Astronomy, The
University of Tennessee, Knoxville, $^{16}$Theory Group, Fermi
National Accelerator Laboratory, $^{17}$Department of Physics,
Boston University, $^{18}$Santa Maria University, Chile,
$^{19}$Department of Physics and Astronomy, University of
Pennsylvania, $^{20}$Department of Physics and Astronomy, University
of Kansas, $^{21}$Department of Physics, University of Maryland,
$^{22}$Department of Physics, Northeastern University,
$^{23}$Department of Physics, University of California, Berkeley and
Lawrence Berkeley National Laboratory, $^{24}$Department of Physics
and Astronomy, University of Minnesota, $^{25}$Standford Linear
Accelerator Center, Standford University, $^{26}$Santa Cruz
Institute for Particle Physics, Department of Physics, University of
California Santa Cruz, $^{27}$Department of Physics and Astronomy, Stony Brook
University, $^{28}$Department of Physics, Oklahoma State University,
 $^{29}$Department of Physics and Astronomy, Iowa
State University, $^{30}$Department of Physics, Carnegie Melon.}
}%

\date{October 10, 2008}

\maketitle

\section{Executive Summary}

The scientific case for a Deep Underground Science and Engineering
Laboratory [DUSEL] located at the Homestake mine in Lead, South
Dakota is exceptional.   The site of this future laboratory already
claims a discovery for the detection of solar neutrinos, leading to
a Nobel Prize for Ray Davis.  Moreover this work provided the first
step to our present understanding of solar neutrino oscillations and
a chink in the armor of the Standard Model of particle physics. We
now know, from several experiments located in deep underground
experimental laboratories around the world, that neutrinos have mass
and even more importantly this mass appears to fit into the
framework of theories which unify all the known forces of nature,
i.e.  the strong, weak, electromagnetic and gravitational.

Similarly, DUSEL can forge forward in the discovery
of new realms of nature, housing {\it six fundamental experiments} that will test the frontiers of our
knowledge:
\begin{enumerate}

\item {\bf Searching for nucleon decay} (the decay of protons and neutrons
predicted by grand unified theories of nature).

\item {\bf Searching for neutrino oscillations and CP violation} by detecting
neutrinos produced at a neutrino source (possibly located
at Brookhaven National Laboratory and/or Fermi National Laboratory).

\item {\bf Searching for astrophysical neutrinos}
originating from the sun, from cosmic rays hitting the upper
atmosphere or from other astrophysical sources, such a supernovae.

\item {\bf Searching for dark
matter particles} (the type of matter which does not interact
electromagnetically, yet provides 24\% of the mass of the Universe).

\item {\bf Looking for the rare process known
as neutrino-less double beta decay} which is predicted by most
theories of neutrino mass and allows two neutrons in a nucleus to
spontaneously change into two protons and two electrons.

\item {\bf Searching for the rare process of neutron- anti-neutron
oscillations},
which would establish violation of baryon number symmetry.

\end{enumerate}

\noindent A large megaton water Cherenkov detector for neutrinos and
nucleon decay, located in DUSEL and roughly 20 times the size of
current detectors, can perform the first three of these experiments.
The last 3 can utilize the unique environment afforded by DUSEL to
perform the most sensitive tests to date.  Any one of these
experiments can greatly increase our knowledge of nature.

The Deep Underground Science and Engineering Laboratory (DUSEL),
with a Large Megaton Size Detector, is desperately needed to address
a set of fundamental issues in particle and astrophysics.
\begin{itemize}

\item Evidence for proton decay would confirm and test grand unified
theories of the four known forces.  It would open a unique window
onto physics at the very highest energy scales.  However it is
important to recognize that any large proton decay detector is a
multi-purpose discovery observatory.   It is also a powerful
neutrino observatory.

\item  Neutrino oscillations involving the transformation of one species
of neutrino into another has been discovered, but there is a whole
world of New Physics lying buried in the neutrinos.  In particular,
it pertains to the question of the violation of a certain symmetry
called CP, which combines the symmetry C, that interchanges matter
and anti-matter, with the symmetry P that reflects spatial
directions. Learning about the violation of this combined symmetry
CP can shed light on the origin of an excess of matter over
anti-matter in the early Universe, which is crucial to the origin of
life and thus to our very existence. In addition we need to know the
ordering of the masses of the three neutrinos and their mixings
accurately. A large size detector with linkage to a long baseline
neutrino facility would enable us to probe into these issues of
great fundamental importance.

\item Finally, observing in detail solar neutrinos, or detecting the neutrinos from a supernova
explosion or other point sources of cosmic neutrinos will greatly
enhance our knowledge of both neutrinos and the physics of the
exotic engines that produce these neutrinos.

\end{itemize}

\noindent To summarize, the discovery potential of such a detector
is high. It could of course be built in modules , each of about 140
kilotons, and at least five such modules would be needed to achieve
the desired goals. Because of its unique multi-purpose value and its
physics significance, such a Large Size Detector at DUSEL, coupled
to a long baseline neutrino beam (which could be provided for
example by Fermilab), would be one of the greatest assets to the
U.S. and the world as a whole. It would greatly complement the
physics that may be learned from the forthcoming Large Hadron
Collider and would thereby enhance the U.S. High Energy Physics
efforts at the very highest energy frontiers.

Any discovery of dark matter, whether underground or at a collider
or best yet, both, will be associated with new physics beyond the
standard model.  In addition, the discovery of dark matter in an
underground detector can provide a unique window on the distribution
of the dominant form of matter in our galaxy and in the Universe.

The searches for neutrino-less double beta decay or neutron
anti-neutron ($n \bar n$) oscillations have long been identified as
sensitive probes of physics beyond the standard model. The
observation of neutrinoless double beta decay and/or $n \bar n$
oscillations will provide unequivocal evidence that these neutral
particles are their own anti-particles. Either discovery would have
major implications for our understanding of nature.

The theory community is excited by the prospect of DUSEL, since this
would create a U.S. center for studies of proton decay, neutrino
oscillations and astrophysics, dark matter, and the very nature of
the neutrino and the neutron and other possible forefront
experiments probing the properties of nature.  In the history of
particle physics, laboratories have always triggered very fruitful
interactions among theorists and between theorists and
experimentalists. Theory workshops could be hosted at DUSEL, as well
as yearly executive summaries of theoretical progress for
experimentalists and of experimental progress for theorists.

Underground laboratories are now operating at the Gran Sasso mine in
Italy, at Kamioka, Japan and the Sudbury Neutrino Observatory in
Canada, and DUSEL-like facilities are being discussed in Japan,
Europe, and India.   Now is the time for the U.S. to take the lead
in this exciting area of physics, since major discoveries, which can
revolutionize our understanding of nature, are expected.  Finally,
in the forthcoming era of large direct detection collaborations, it
will be strategic for the US scientific community to have a
laboratory like DUSEL in the US, preventing the risk of a drain of
human resources outside the country, with dangerous consequences for
both our experimental and theoretical physics community.  Also, the
health of our economy depends on the education of our future
scientists and engineers and the technology they can develop. A
national facility such as DUSEL can be a guiding light for our
future scientists and engineers.

\newpage

\section{Search for Proton Decay at a Megaton Observatory }

\subsection{Executive Summary}

The search for proton decay at a Large Size Detector can
dramatically shed light on the fundamental aspects of the laws of
nature. In particular:
\begin{itemize}
\item Improved studies of proton decay would enable us to probe
nature at the highest energy scale of order $10^{16}$ GeV (which is
a trillion times larger than the energy that would be available at
the LHC), or equivalently at truly short distances of
order $10^{-30}$ cm---something that would not be possible by any
other means.

\item  The discovery of proton decay would have profound significance
for unification ideas.  The idea of grand unification proposes, on
aesthetic grounds, to unify the basic constituents of nucleons
(so-called quarks) and the non-nuclear particles, like electrons and
neutrinos, as aspects of one kind of matter. Simultaneously it
proposes a unity of the three basic forces---the strong, weak and
electromagnetic. This idea predicts, contrary to the belief commonly
held till the 1970s, that the proton must decay,  albeit with a long
lifetime exceeding $10^{30}$ years. While proton decay has yet to be
seen, the grand unification idea has turned out to be spectacularly
successful as regards its other predictions. These include in
particular the phenomena of ``coupling unification," amounting to an
equality of the strengths of the three forces at very high energies,
which has been verified to hold at an energy scale of $10^{16}$ GeV
by the precision measurements carried out at the CERN Laboratory in
Geneva in the 1990s. Furthermore, a class of grand unified models
naturally predict that the heaviest of the three neutrinos should
have a mass in the range of a hundredth to one electron-Volt, and
the next-to-heaviest an order of magnitude lighter, the two being
quantum-mechanical mixtures of what one calls nu-mu and nu-tau. This
too is in full accord with the discovery of neutrino oscillation at
the Super-Kamiokande Laboratory in Japan in 1998. In this sense,
proton decay now remains as {\em THE MISSING PIECE} of grand
unification.

With the discoveries of both coupling unification, at the scale of
$10^{16}$ GeV, and of neutrino oscillation, one can in fact argue,
within a class of well-motivated ideas on grand unification, that
proton decay should occur at accessible rates, with a lifetime of
about $10^{35}$ years, within a factor of ten either way,  for
protons decaying into positron plus neutral pion, and a lifetime of
less than a few $\times 10^{34}$ years [in theories with 3 space
dimensions] for proton decaying into anti-neutrino + a positively
charged K-meson. Moreover, it is a very exciting fact that whether
the former or latter decay mode dominates depends on the number of
space dimensions.  With 3 spatial dimensions the latter mode
typically dominates, while in higher dimensions, the former modes
can give lifetimes of order $10^{34}$ to $10^{35}$ years. Hence,
these predictions lie at a striking distance---within a factor of
about 5 to 10---above the current lower bound set by the
Super-Kamiokande Laboratory. Thus, unless the successes listed above
are mere coincidence, there is a strong likelihood that proton decay
would be discovered, IF one can improve the current sensitivity (of
Super-Kamiokande) by a factor of 5 to 10. This is why an improved
search for proton decay, possible only with a Large Size Detector,
is now most pressing. Proton decay, if found, would no doubt
constitute a landmark discovery for mankind.

\end{itemize}

\subsection{Proton Decay}

Protons, neutrons and electrons are the fundamental building blocks
of all stable matter.   They are the basic ingredients for chemistry
and biology.   Neutrons are stable when found deep inside the
nucleus of an atom.   However free neutrons are known to decay. When
a neutron decays (there one minute, gone the next) it is replaced by
a proton, an electron and the mysterious particle, called a
neutrino.   Neutrons and protons are held tightly inside the nucleus
via strong nuclear forces.   Neutrons decay via the weak force. The
neutron lives on average about 1000 seconds when free.  This number
is called the lifetime of the neutron and is given in terms of the
relation $\tau_{neutron} \sim  \frac{M_W^4}{\Delta m_N^5}$, where
$\Delta m_N \approx m_n - m_p$ is the neutron - proton mass
difference and $M_W$ is the $W$ boson's mass.   The weak
interactions are so weak because the $W$ boson is much heavier than
the nucleon (proton or neutron) and  $\frac{M_W}{\Delta m_N} \approx
30,000$.

But is the proton stable?   If it could decay, then what would it
decay into?   What would its lifetime be?   When a neutron decays,
it does so preserving a quantity called {\em baryon number} or {\em
baryon charge}. Both protons and neutrons interact via strong
nuclear forces and both are {\em baryons} with {\em baryon charge $B
= +1$}. The anti-particles of the proton and neutron exist and they
have {\em baryon charge $B = -1$}. When a proton (with $B = +1$) and
an anti-proton (with $B = -1$) meet they annihilate releasing their
considerable mass energy ({\em recall $E = m c^2$}) into other forms
of energy with total $B = 0$. On the other hand, when a neutron ($B=
+1$) decays, it decays into a proton (also with $B = +1$) and an
electron and anti-neutrino (both with $B = 0$).  The sum of the
baryon charges of the initial and final states agree!  Hence baryon
charge is conserved. Protons are the lightest baryons.  All observed
processes appear to conserve baryon charge.  For a proton to decay
it must conserve energy and electric charge.  Electric charge and
energy conservation laws are associated with long range forces, i.e.
electromagnetic and gravitational, respectively.   Baryon charge
conservation, on the other hand, is NOT associated with any long
range force.  In fact, not only is it not understood why baryon
charge should be conserved, most theories beyond the standard model
do NOT conserve baryon charge.

If baryon charge conservation were violated then protons might
decay, for example via the processes $p \rightarrow e^+ \pi^0$ or $p
\rightarrow  e^+ \gamma$, where $e^+$ is the anti-electron (or
positron), $\pi^0$ is the neutral pi meson and $\gamma$ is a photon.
If protons were to decay rapidly, then all chemistry and life as we
know it would come to an abrupt end. The fact that life exists at
all implies $\tau_p > 10^{18}$ years, since each and every proton or
neutron (eg. $n \rightarrow \bar \nu + \pi^0$) in our body can decay
and release its mass in deadly radiation. Similar to the weak decay
of the neutron, the proton lifetime is given by an expression of the
form $\tau_p \sim M^4/m_p^5$ where $M$ is a new scale of nature. The
existence of life then implies $M > 10^{12}\; m_p$. In units
appropriate to accelerator energies, the proton mass, $m_p \approx
10^9 \; {\rm electron- Volt}/c^2 = 1$ GeV$/c^2$ (Giga-eV$/c^2$). For
comparison, the Fermilab Tevatron has a maximum energy of 1.8
$\times 10^{12}$ eV = 1.8 TeV (Tera-eV), while the Large Hadron
Collider, soon to turn on at CERN in Geneva, Switzerland, will have
an energy of 14 TeV. The bottom-line is that the search for proton
decay explores physics at the highest energies, much higher than is
reachable in any accelerator experiment. Moreover, as we discuss
later, an observation of proton decay would forever change our
understanding of nature, with ramifications for understanding why
there is more matter, than anti-matter in the Universe,  and the
proposed grand unification of the four known forces of nature.

\subsection{Matter- anti-Matter asymmetry of the Universe}

We are made of baryons,  NOT anti-baryons.  But why is this so? The
answer: baryon charge must be violated. Without baryon charge
violation (assuming equal numbers of baryons and anti-baryons
initially)  it is easy to show that most of the baryons and
anti-baryons in the Universe would annihilate and we would be left
with too few baryons.  Why assume equal numbers? Otherwise, we would
require the baryon to anti-baryon asymmetry to be set by (some)
hand.   This said, it has been shown that baryon charge violating
processes can be used to derive the observed matter- anti-matter
asymmetry.

\subsection{Probing high energies with proton decay at a Large Size Detector}

Any new physics {\em beyond the standard model} predicts new energy
scales with new particles and forces.  In many cases,  baryon charge
violating interactions are also expected.   One well-motivated idea
for new physics beyond the standard model is known as {\em
supersymmetric grand unified theory}.  Grand unification describes
the unity of the strong, weak and electromagnetic interactions, as
well as the the unity of quarks and leptons.  Grand unified theories
are also naturally incorporated into {\em superstring theory}
resulting in the unification of strong, weak, electromagnetic {\em
and gravitational} interactions; the penultimate unification!

We expect the grand unification of the strong nuclear force with the
weak and electromagnetic forces at a scale $M_G \sim 10^{16}$ GeV.
Grand unification of strong, weak and electromagnetic interactions
requires unification not only of these forces but it also requires
the unification of quarks and leptons.  Meaning that quarks and
leptons are indistinguishable at their most fundamental level.  This
is not just the hope of enthusiastic physicists,  it is in fact
suggested by data.  The LEP experiment at CERN made the most precise
measurements of the coupling strengths of the strong, weak and
electromagnetic interactions.  Using this data it was shown that the
these three couplings can unify at a grand unification scale $M_G$,
provided that there is a doubling of the particles in nature, i.e.
provided that this doubling is described by {\em supersymmetric
GUTs}. Thus IF grand unification is real, then these new
supersymmetric partners of ordinary matter should be observable at
the Large Hadron Collider soon to take data at CERN!! Thus the LHC may
open another unique window onto energy scales of order the Planck
scale (where gravity becomes strong).

Independent of grand unification the new scale $M$ of order
$10^{14}$ GeV can be used to explain the observation of neutrino
oscillations. Electron and muon neutrinos (there are three families
of leptons, called electron, muon and tau, each with its own
neutrino species) are produced in the upper atmosphere when cosmic
ray protons hit air molecules. It has been demonstrated by
experiments at Super-Kamiokande in Japan that muon neutrinos change
into tau neutrinos on their way to the surface of the earth. This
experimental result is explained by neutrino mass. However neutrinos
are nevertheless some 100 million times lighter than the electron.
This amazing fact is explained most naturally by the so-called
See-Saw mechanism.   The light neutrino is {\em light} as a
consequence of a very {\em heavy} new energy scale $M \sim 10^{10}
\; {\rm to} \; 10^{14}$ GeV, with $m_{\rm neutrino} \sim (m_{\rm
lepton})^2/M$.   Finally neutrino mass and neutrino oscillations are
a natural component of grand unified theories.

Proton decay with accessible rates is a crucial prediction of the
idea of grand unification (see Fig. 1).  What is the expected
lifetime of the proton?  The answer to this question depends on
whether the theory is realized in four space-time dimensions or in
higher dimensions, as might be expected in string theory,  such as
in 5, 6 or 10 dimensions with sizes not much larger than the Planck
length. In four dimensions, very conservatively, most models predict
an upper bound on the proton lifetime $\tau(p \rightarrow K^+ \bar
\nu) \leq  \; {\rm few} \; \times 10^{34}$ yrs., while for the other
dominant decay mode, $\tau(p \rightarrow e^+ \pi^0) \sim {\cal O}
(10^{36}$ years). On the other hand, for grand unified theories in
higher dimensions {\em $\tau(p \rightarrow e^+ \pi^0)$ can be as low
as $10^{33}$ years.}

      \begin{figure}
\scalebox{0.6}
      { \includegraphics[width=20cm,height=16cm]{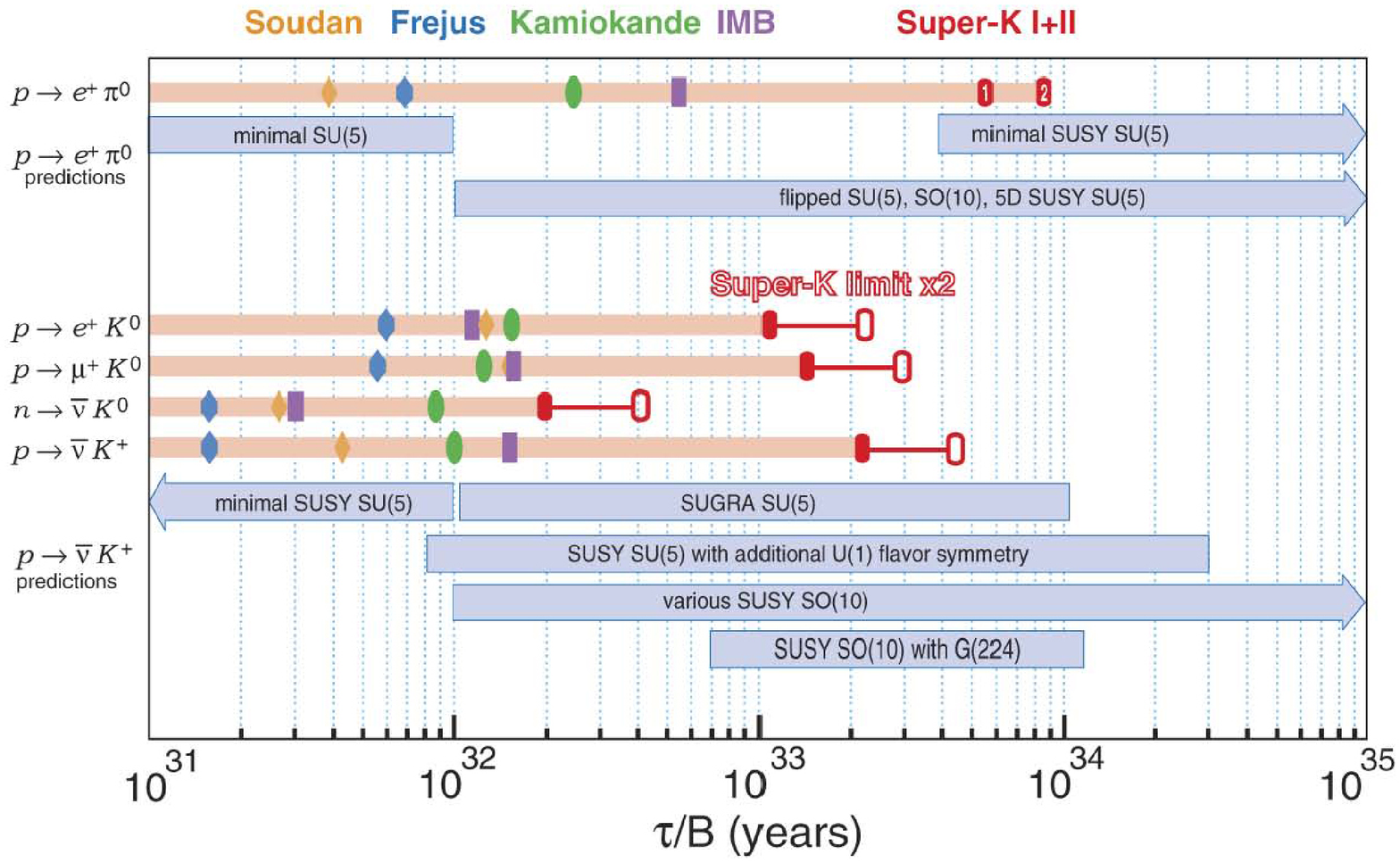}
%~\\~\\
      }
      ~\\~\\
\caption{Super-Kamiokande bounds on some prominent proton and
neutron decay channels with predictions
 from some well-motivated theories.  Courtesy of E. Kearns, NNN07 talk.  }
      \label{fig:guts}        \end{figure}

Perhaps grand unification is wrong. Perhaps the LHC will explore
large extra dimensions with strong gravitational interactions as low
in energy as $M \sim 10$ TeV.  Such theories are strongly
constrained by the non-observation of proton decay. They must
naturally have, or impose, a symmetry strongly suppressing baryon
charge violation. For example, if space-time is 6 dimensional ({\em
with two large extra dimensions}), then it was shown that the proton
decays via the process $p \rightarrow e^- \pi^+ \pi^+ \nu \nu$. This
decay is highly suppressed and leads to a lifetime on the order of
$10^{35}$ years. However, in more general large extra dimension
models an almost exact baryon charge symmetry must be imposed or
else they are already excluded by the non-observation of proton
decay.

Finally, strong gravitational processes are expected to induce
proton decay at a scale $M \sim M_{Pl} = 10^{19}$  GeV, where
$M_{Pl}$ is the scale at which Newtonian gravity becomes strong.
This would give $\tau_p > 10^{46}$ years, (assuming nature is NOT
supersymmetric) which would be unobservable by all proposed proton
decay experiments. But this dour prediction is an unlikely scenario.
In addition, if supersymmetry is discovered at the LHC, then even if
there is no grand unification, we still expect proton decay at
observable rates just from strong gravitational induced processes.

To summarize, {\em every grand unified theory predicts that the
proton will decay}.   Moreover, we may need to describe nature (with
or without grand unification) in more than three space dimensions or
eventually with superstring theory.  Indeed, in every single case it
is expected that the proton will decay.  The dominant decay mode for
the proton (and neutron) is model dependent.  Yet the lifetime is
typically less than $10^{36}$ years, {\em and in many cases it is
much lower}.

\subsection{History of proton decay experiments}

The dedicated search for proton decay began in the early 80s.  The
best bounds now come from the Frejus experiment, France; Soudan 2
and IMB, USA, and Kamiokande (and Super-Kamiokande), Japan.
These experiments have not seen evidence for proton
decay.  Thus they set limits on the proton lifetime. The best limit
from Super-Kamiokande (preliminary) is $\tau(p \rightarrow e^+
\pi^0) > 8.4 \times 10^{33}$ yrs or Super-Kamiokande I for the mode
$\tau(p \rightarrow K^+  \bar \nu)
> 2.3 \times 10^{33}$ yrs.

Nevertheless, notwithstanding the lack of proton decay events, the
IMB experiment (located near Cleveland, OH) and Kamiokande made a
stupendous, serendipitous discovery of super-Nova neutrinos in 1987;
confirming the theory of super-nova collapse!! In addition,
Super-Kamiokande has been key to understanding solar and atmospheric
neutrino oscillations. Hence an experimental program started
initially to see protons decay has been instrumental in our
understanding of neutrino properties, as well as our understanding
of astrophysics; {\em two major successes}.

\subsection{Future proton decay detectors}

A large size water Cherenkov detector is proposed for the DUSEL
site. It is a megaton detector which can be built in 140 kiloton
modules. With 5 such modules,  the proton decay measurements can be
sensitive to a lifetime of order $10^{35}$ years (see Fig.s 2 and
3); good enough to test most models for new physics beyond the
standard model of particle physics.  It is important to note that
any proton decay detector is also a neutrino observatory. All such
detectors are designed with a multi-prong experimental program which
includes,
\begin{itemize} \item proton decay, \item long baseline neutrino oscillations to
measure the remaining unmeasured mixing angle, $\theta_{13}$, and CP
violation in the lepton sector, and
\item the observation of astrophysical neutrinos, such as a near-by
super-nova or relics of past super-novae. \end{itemize}

Several large detectors, in different parts of the world, have also
been proposed to continue the search for proton decay. These
include, Hyper-Kamiokande in Japan and LAGUNA in Europe.
Hyper-Kamiokande is a water Cherenkov detector, while LAGUNA is a
European collaboration which is considering three possible
technologies; water Cherenkov, liquid argon or liquid scintillators.
These detectors, if built, have similar goals to any DUSEL detector,
i.e. to reach a lifetime sensitivity of $10^{35}$ years.

      \begin{figure}
\scalebox{0.6}
      { \includegraphics[width=20cm,height=16cm]{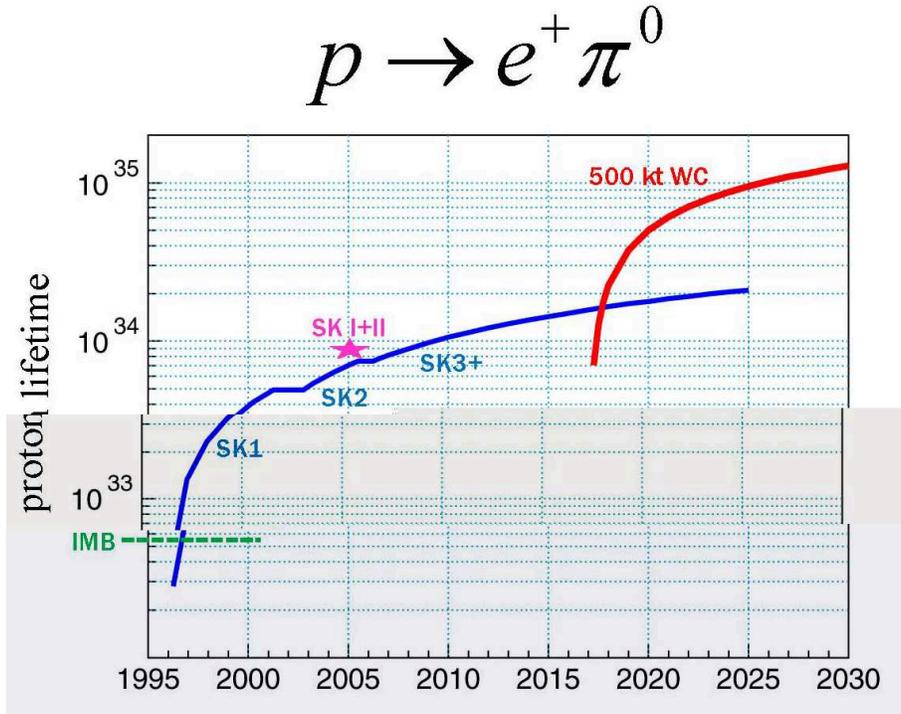}
%~\\~\\
      }
      ~\\~\\
 \caption{The blue line gives the present \& future Super-K bounds (as a function of time) for the proton lifetime
 into the decay mode $p \rightarrow e^+ \pi^0$. The red line indicates the advantage of a half Megaton water Cherenkov
 detector.  Courtesy of E. Kearns, NNN07 talk. }
      \label{fig:pepi0}        \end{figure}

      \begin{figure}
\scalebox{0.6}
      { \includegraphics[width=20cm,height=16cm]{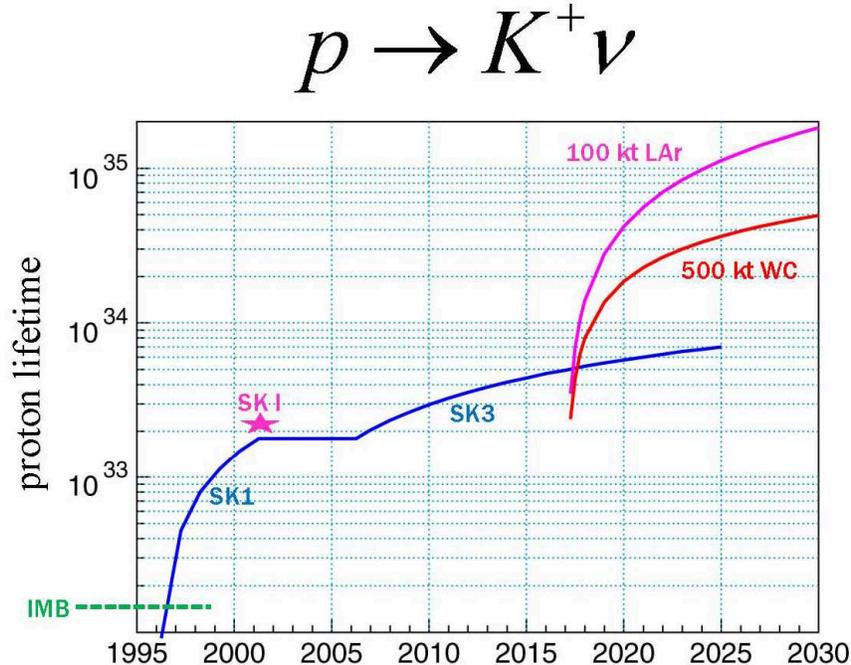}
%~\\~\\
      }
      ~\\~\\
\caption{The blue line gives the present \& future Super-K bounds
(as a function of time) for the proton lifetime
 into the decay mode $p \rightarrow K^+ \nu$. The red (or magenta) line
 indicates the advantage of a 1/2 Megaton water Cherenkov
 (or 100 kton liquid argon) detector.  Courtesy of E. Kearns, NNN07 talk. }
      \label{fig:pknu}        \end{figure}

It is thus crucial that the U.S. is a competitor in this super high
energy frontier.  This is not only to achieve the awesome science
goals, but also to reap the benefits to our educational system and
our culture that this search ({\em and discovery}) will bring. For
this to be possible, we must start as soon as possible to construct
a Megaton Observatory for Neutrinos and for Nucleon Decay.

\subsection{Summary}

The Deep Underground Science and Engineering Laboratory (DUSEL),
with a Large Megaton Size Detector, is desperately needed to address
a set of fundamental issues in particle and astrophysics. The
discovery potential of such a detector is high. It could of course
be built in modules , each of about 140 kilotons. At least five such
modules would be needed to achieve the desired goal.  Because of its
unique multi-purpose value and its physics significance, such a
Large Size Detector at Homestake, coupled to a long baseline
neutrino beam (which could be provided for example by Fermilab),
would be one of the greatest assets to the U.S. and the world as a
whole. It would greatly complement the physics that may be learned
from the forthcoming Large Hadron Collider and would thereby enhance
the U.S. High Energy Physics efforts at the very highest energy
frontiers.

\newpage

\section{Long Baseline Neutrino Experiment}

\subsection{Executive Summary}

A continuing program designed to study CP violation in the neutrino
sector and to determine the hierarchy of the neutrino mass spectrum
is scientifically compelling.  The US program may be unique in the
world in its ability to measure the ordering of the neutrino mass
spectrum.

CP violation has so far only been observed in the quark sector of
the standard model.  Its discovery in the neutrino sector should
shed additional light on the role of CP violation in nature.
Unveiling neutrino CP violation is particularly important because of
its potential connection with the observed matter--antimatter
asymmetry of our Universe, a fundamental problem at the heart of our
existence. The leading explanation is currently a leptogenesis
scenario in which decays of very heavy right--hand neutrinos created
in the early Universe give rise to a lepton number asymmetry which
later becomes a baryon--antibaryon asymmetry. Leptogenesis offers an
elegant, natural explanation for the matter--antimatter asymmetry;
but it requires some experimental confirmation of its various
components before it can be accepted. Those include the existence of
very heavy right--handed neutrinos as well as lepton number and CP
violation in their decays.

A number of neutrino mass models have been proposed and precise
knowledge of neutrino parameters is essential to test them.
Specifically, the value of the mixing angle $\theta_{13}$ and the
hierarchy of the neutrino mass spectrum will help distinguish
between models based on lepton flavor symmetries, models with
sequential right-handed neutrino dominance and more ambitious models
based on Grand Unified Theory (GUT) symmetries.  GUT models
naturally yield a particular ordering of the neutrino spectrum and a
relatively large $\theta_{13}$.

The wide-band beam approach has a greater scientific reach for
neutrino oscillations when located at a distance that permits
resolution of the neutrino mass hierarchy, and further scope if the
detector is located at a depth that permits the study of nucleon
decay.

\begin{itemize}

\item The scientific goals of a program of long baseline neutrino
oscillation experiments are to measure the mixing parameter
$\sin^22\theta_{13}$, to determine the order of the states of the
neutrino mass spectrum, and to determine whether there is CP
violation in the neutrino sector.  Measurement of these quantities
is an important goal of elementary particle physics.

\item Determination of the ordering of the neutrino mass spectrum,
searching for CP violation, and resolution of parameter degeneracies
with sensitivity down to $\sin^22\theta_{13}\simeq 0.01$ will
require a new generation of experiments with detectors with mass of
100 kilotons or more. This represents an increase in sensitivity of
more than one order of magnitude over the experiments that will
begin to acquire data in the next few years.

\item The wide-band beam approach to neutrino oscillation physics can,
in principle, utilize either a liquid argon detector or a water
Cherenkov detector.  If located more than 1000 km from Fermilab,
there is good sensitivity for determining the mass hierarchy and
measuring the amount of $CP$ violation. The optimal baseline for a
wide band beam experiment is between 1200 and 1500 km.

\item Among experiments with super neutrino beams, wide band beam experiments have the most robust performance and the best mass hierarchy performance. Overall, they are the optimal choice to pursue after the near-future reactor and narrow band beam experiments.

\end{itemize}

\subsection{Report on Neutrino Oscillations}

Neutrinos, nearly massless and electrically neutral elementary
particles, provide a unique window on the structure of matter at
subatomic scales.  They exist in three types: electron, muon and
tau. In the past decade muon-neutrinos produced in cosmic ray
reactions in the earth's atmosphere and electron-neutrinos produced
in nuclear reactions in the sun's core have been shown to change
from one kind to another between their source and detection. Further
experimentation with both natural neutrino sources and neutrinos
from reactors and accelerators has shown that the quantum mechanical
mixing of neutrino types, also known as neutrino oscillations, is
responsible for this change.  A new generation of experiments has
been initiated using reactor and accelerator neutrinos to make
precise measurements of the mixing phenomena.

The discovery of neutrino oscillations showed that neutrinos have
mass that are a million times smaller than the mass of the next
lightest elementary particle, the electron.  The reason the neutrino
masses are so small is a fundamental issue that must be understood.
It is expected that physics at energies much higher than those
available in our laboratories are responsible for the origin of
neutrino mass. Neutrinos are so abundant that the total mass of all
the neutrinos in the Universe may be comparable to the total mass of
all the stars in the Universe.  Continuing studies of neutrinos will
illuminate the most basic issues in physics at very small distance
scales and at very large distance scales.

The three observed neutrino types, called flavors, couple to other
particles with strengths given by the standard model of elementary
particles. Th quantum mechanical mixing is parameterized by three
mixing angles, $\theta_{12}$, $\theta_{23}$ and $\theta_{13}$, and
one phase angle, $\delta_{CP}$, the so-called $CP$ phase. The angle
$\delta_{CP}$ describes how neutrinos and antineutrinos differ in
their interactions with matter.

Neutrino oscillation phenomena depend on the four angles and the
difference in the squares of masses ($\Delta m^2$) of the
participating neutrinos.  The discovery of atmospheric neutrino
oscillations in the Super-Kamiokande experiment demonstrated that
$\Delta m^2_{32} \simeq \pm2.5\times 10^{-3} {\rm eV}^2$ and mixing
angle $\theta_{23}\simeq 45^{\circ}$. These findings have been
confirmed and made more precise by the MINOS experiment with an
accelerator generated neutrino beam from Fermilab directed at a
detector in the Soudan mine in Minnesota.

As yet, the sign of $\Delta m^2_{32}$ is undetermined.  The
so-called normal mass hierarchy, $m_1, m_2 < m_3$, suggests a
positive sign which is often obtained in theoretical models.
However, a negative value (or inverted hierarchy, $m_1, m_2 > m_3$)
can certainly be accommodated, and if that is the case, the
predicted rates for neutrinoless double beta decay will likely be
larger and more easily accessible experimentally.  Resolving the
sign of the mass hierarchy is an extremely important issue.  In
addition, the fact that $\theta_{23}$ is large and near maximal is
also significant for model building.  Measuring that parameter with
precision is highly desirable.

The deficit of observed neutrinos from the sun compared to
expectations was a decades-long puzzle that has been definitively
explained as due to oscillations of solar neutrinos as they
propagate through the sun.  From measurements of solar and reactor
neutrino oscillations it has been found that $\Delta m^2_{21} \simeq
8\times 10^{-5} {\rm eV}^2 $ and $\theta_{12}\simeq 32^{\circ}$. The
sign of $\Delta m^2_{21}$ is known to be positive due to the effects
of the solar medium on the propagation.

The mixing angles $\theta_{12}$ and $\theta_{23}$ are large relative
to all of the mixing angles in the quark sector. The reason for the
different patterns of mixing in the neutrino and quark sectors
remains to be understood theoretically.  In addition, $\Delta
m^2_{21}$ is large enough, compared, to $\Delta m^2_{32}$, to make
long baseline neutrino oscillation searches for CP violation
feasible and could yield positive results, i.e. the stage is set for
a future major discovery of CP violation in the lepton sector.

\begin{figure}
\includegraphics*[scale=0.5]{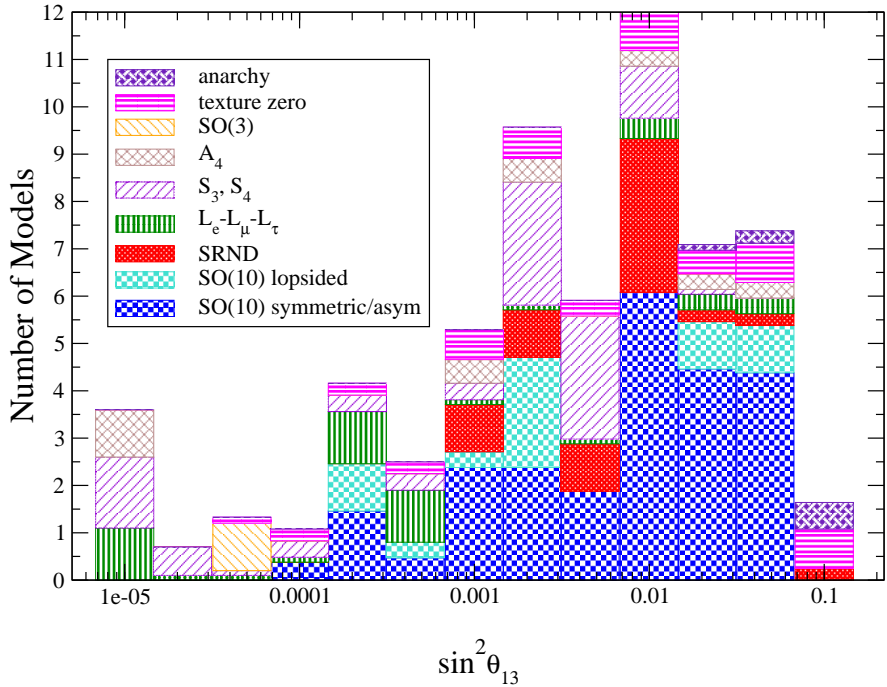}
\caption{Histogram of $\sin^2 \theta_{13}$ predictions for 63
models.  Source: C.H. Albright and M.C. Chen,
arXiv:hep-ph/0608137.} \label{fig:hist}
\end{figure}

Currently, we know nothing about the value of the $\deltacp$ and
only have an upper bound on the as yet unknown mixing angle
$\theta_{13}$ ($\theta_{13} < 13^{\circ}$ or $\sin^22\theta_{13}\leq
0.2).$ However, a survey of 63 models in the literature found that
the predictions for $\theta_{13}$ were clustered around
$\sin^22\theta_{13} = 0.04$ ($\sin^2\theta_{13} = 0.01$), as
displayed in Fig.~\ref{fig:hist}. If $\sin^22\theta_{13}$ is
comparable to or greater than this value, it is likely to be
determined by the coming generation of reactor $\bar\nu_e$
disappearance experiments at Double CHOOZ (France) and Daya Bay China)
and the upcoming accelerator based $nu_{\mu}\to\nu_e$ appearance
 experiments T2K (Japan, J-Parc to Super-Kamiokande) andNO$\nu$A (USA,
 Fermilab to Minnesota).  Reactors experiments are complimentary to long-baseline
experiments in that they can provide valuable information on
$\theta_{13}$ but not on the mass hierarchy or $\deltacp$.

Based on our current knowledge and future goals, a future neutrino
program should include the following objectives:

\begin{itemize}

\item Complete the measurement of the neutrino mixing angles,

\item Determine the sign of $\Delta m^2_{32}$,

\item Measure $\deltacp$ to determine if CP is violated,

\item Search for exotic effects in neutrino oscillations.

\end{itemize}

\medskip
\noindent Of the above future neutrino physics goals, the search for
and study of CP violation is of primary importance and should be our
main objective for several reasons, which we briefly address.

CP violation has so far only been observed in the quark sector of
the standard model.  Its discovery in the neutrino sector should
shed additional light on the role of CP violation in nature.
Unveiling neutrino CP violation is particularly important because of
its potential connection with the observed matter--antimatter
asymmetry of our Universe, a fundamental problem at the heart of our
existence. The leading explanation is currently a leptogenesis
scenario in which decays of very heavy right--hand neutrinos created
in the early Universe give rise to a lepton number asymmetry which
later becomes a baryon--antibaryon asymmetry via the B-L conserving
't Hooft mechanism of the standard model at weak scale temperatures.

Leptogenesis offers an elegant, natural explanation for the
matter--antimatter asymmetry; but it requires some experimental
confirmation of its various components before it can be accepted.
Those include the existence of very heavy right--handed neutrinos as
well as lepton number and CP violation in their decays.

A number of neutrino mass models have been proposed and precise
knowledge of neutrino parameters is essential to test them.
Specifically, the value of the mixing angle $\theta_{13}$ and
whether the mass hierarchy is normal or inverted will help
distinguish between models based on lepton flavor symmetries, models
with sequential right-handed neutrino dominance and more ambitious
models based on Grand Unified Theory (GUT) symmetries.  GUT models
naturally yield a normal hierarchy and a relatively large
$\theta_{13}$ (although in a few unified models, an inverted
hierarchy can be obtained with fine-tuning).

Leptogenesis can naturally emerge in grand unified theories.
Moreover, successful unification of the strong, weak and
electromagnetic forces strongly suggests the existence of a
supersymmetry and its associated new particles at the TeV scale. The
lightest stable particle of supersymmetry is a leading candidate for
dark matter in the Universe. Thus neutrino physics is intimately
connected to the most interesting outstanding questions today that
are to be explored at the Large Hadron Collider, dark matter
detection experiments and the IceCube experiment (which is designed
to look for very high energy neutrinos as they pass through the
Antarctic ice cap).

Designing for CP violation studies in next generation neutrino
programs has other important benefits.  First, the degree of
difficulty to establish CP violation is high but achievable.  It
requires an intense proton beam of about 1--2~MW and a very large
detector, $100$-$500$ kton Water Cherenkov (WC) or a liquid argon
(LArTPC) detector of size $\sim 100$ kTon which could be equivalent
in sensitivity due to its better performance. Water Cherenkov is an
established technology, while liquid argon, which promises superior
particle identification and control over backgrounds, is still under
development.  Such an ambitious infrastructure will allow very
precise measurements of all neutrino oscillation parameters as well
as the mass hierarchy via $\nu_{\mu}\to \nu_{\mu}$ disappearance and
$\nu_{\mu}\to \nu_{e}$ appearance studies.

\begin{figure}[tp!]
\begin{center}
\includegraphics[height=18cm]{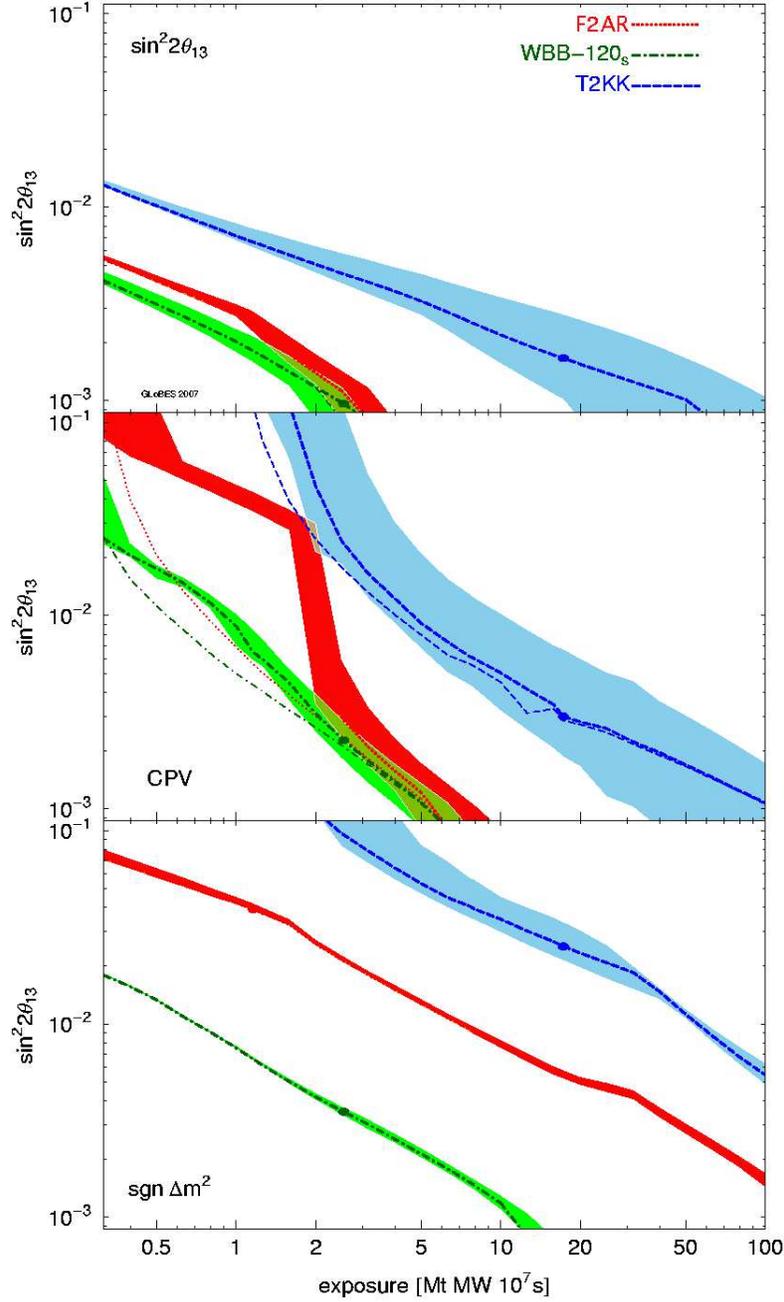}
\end{center}
\caption{The $\sin^22\theta_{13}$ reach at $3\sigma$ for the
discovery of nonzero $\sin^22\theta_{13}$, CP violation, and the
normal hierarchy as a function of exposure. The curves are for a
fraction of $\deltacp$ of 0.5, which means that the performance will
be better for 50\% of all values of $\deltacp$, and worse for the
other 50\%. The light curves in the CPV panel are made under the
assumption that the mass hierarchy is known to be normal. The dots
mark the exposures of the setups as defined in
Table~\ref{tab:setups}. The shaded regions result by varying the
systematic uncertainties from 2\% (lower edge) to 10\% (upper edge).
} \label{fig:lumiscale}
\end{figure}

In order to make an unbiased comparison of the physics potentials of
the experimental setups the sensitivities as functions of {\it
{exposure}} may be compared, where exposure is defined to be
$\mathcal{L}=$ detector mass [Mt] $\times$ target power [MW]
$\times$ running time [$10^7$~s]. The relative merits of three
superbeam scenarios, beta beam ($\beta$-beam) experiment, and
neutrino factory (NuFact) experiment, are listed in
Table~\ref{tab:setups}.  In Fig.~\ref{fig:lumiscale} we show the
discovery reaches for $\sin^22\theta_{13}$, CP violation, and normal
mass hierarchy versus the exposure for a fraction of $\deltacp$ of
0.5 (see figure caption). The experiments we considered are a future
narrow-band beam experiment from Fermilab to Ash River (F2AR) with
average neutrino energy $E_\nu = 2.6$~GeV, a wide-band beam
experiment ($E_\nu = 2.6$), and a narrow-band beam experiment from
Tokai to Kamioka and Korea (T2KK) with $E_\nu \simeq 0.8$~GeV.

\begin{table}
{\small
\begin{tabular}{lcccccr}
\hline Setup & $t_{\nu}+t_{\bar\nu}$ [yr] & $P_\mathrm{Target}$ [MW]
& $L$ [km] & Detector technology &
$m_{\mathrm{Det}}$ [kt] & $\mathcal{L}$ \\
\hline F2AR & 3 + 3 & 1.13 ($\nu$/$\bar{\nu}$) & 810 &
LArTPC & 100 & 1.15 \\
WBB &  5 +  5 & 1 ($\nu$) +2
($\bar{\nu})$& 1290 & LArTPC & 100 & 2.55 \\
T2KK &  4 + 4 & 4 &
295+1050 & WC & 270+270 & 17.28 \\
\hline
$\beta$-beam & 4 + 4 & n/a & 730 & WC & 500 & n/a \\
NuFact &  4 + 4 & 4 & 3000+7500 & Magn. iron calor. & 50+50 & n/a
\\
\hline
\end{tabular}
} % small
\caption{Setups considered, neutrino $t_{\nu}$ and antineutrino
$t_{\bar{\nu}}$ running times, corresponding target power
$P_\mathrm{Target}$, baseline $L$, detector technology, detector
mass $m_{\mathrm{Det}}$, and exposure $\mathcal{L}$ [$\mathrm{Mt \,
MW \, 10^7 \, s}$].} \label{tab:setups}
\end{table}

If $\theta_{13}$ is not too small ($\sin^22\theta_{13} \simeq
0.01$), it may be possible to mount experiments that will permit us
to determine the ordering of the states in the neutrino mass
spectrum and to measure CP violation in the neutrino sector of the
particle world. To pursue this goal a wide-band, on-axis, neutrino
beam directed to a future Deep Underground Science and Engineering
Laboratory site, needs to be developed.  The fact that a
very large underground detector can also be used to determine
neutrino CP violation and measure all facets of neutrino
oscillations gives such a facility an outstanding discovery
potential.

\newpage

\section{Astrophysical Neutrinos}
\subsection{Executive Summary}

Until the 1950's visible light provided our only view of the
Universe.  Tremendous progress came in the following years when we
extended our observational toolbox to the whole electromagnetic
spectrum from microwaves to gamma rays. Recently a new window on the
Universe has opened---neutrino astronomy and astrophysics---and
there is hope that observing cosmological and astrophysical
neutrinos, some from the most exciting electromagnetic sources in
the Universe, will likewise expand our understanding. We are now
able to detect neutrinos of astronomical origin with energies
ranging from a few MeV ($10^{6}$ eV) to $10^{21}$ eV. The lower end
of this energy range is accessible at underground laboratories
whereas the higher end is accessible with huge neutrino telescopes
such as IceCube or ANTARES, which use the Antarctic ice cap or the
Mediterranean Sea, respectively. Astrophysical neutrinos help us
explore fascinating phenomena in the Cosmos as diverse as the birth
of new stars and the origin of elements. These neutrinos provide a
new tool complementary to other tools already in place: various
electromagnetic (optical or otherwise) telescopes, which can look at
the photons coming from neutrino-emitting objects, or LIGO, which
can measure gravitational collapse accompanying neutrinos. We have
already seen neutrinos from two (and only two) such objects: A
main-sequence star (the Sun) and a core-collapse supernova
(SN1987A). In both cases, deep and fundamental understanding about
the nature of the Universe resulted.

It should be emphasized that there is a large and active community
of researchers in the U.S. to use DUSEL carrying out a program in
this new neutrino astrophysics. Such an activity would be
complementary to the very high-energy neutrino astronomy programs
already in place with experiments such as IceCube.  Recent
accomplishments in this area include:
\begin{itemize}
\item    Theoretical prediction of the solar neutrino flux and
structure  of the main sequence stars are confirmed by solar neutrino measurements,
resulting in the  Standard Solar Model.
For example, the temperature at the
center of the Sun was correctly calculated ab initio to better than
2\%.
\item    Recognition of the importance of the neutrino-neutrino
interactions on neutrino propagation in dense neutrino systems and
the development of the theoretical tools to treat these effects in
astrophysical sites. Thus core-collapse supernovae provide us with
the only example of a non-trivial many-body system entirely
controlled by weak interactions.

\item New theoretical breakthroughs
in understanding nucleosynthesis in supernovae and gamma-ray bursts
as well as the role of weak interactions in supernova dynamics.

\item Establishing that active neutrinos cannot be the dark matter, which
is independently confirmed by cosmological data.
\item  Placing new limits on the diffuse supernova neutrino
background. Reduced astrophysical uncertainties mean that these
searches are primarily testing the neutrino emission per supernova.
In fact, considering the significantly lower reactor neutrino and
cosmic ray backgrounds at the 4850 ft. level of DUSEL compared to
Super-Kamiokande and the use of Gadolinium, there is a strong
likelihood of actually observing the diffuse supernova neutrino
background.
\end{itemize}

\subsection{Report on Neutrino Astrophysics}

The first example of neutrino astrophysics concerned neutrinos from
the Sun. The solar neutrino program represents one of the great
triumphs of physical science in the last 40 years and led to the
Nobel Prize for Ray Davis in 2002. The story starts with an old
question as to the origin of the sun's energy. The answer analyzed
by Hans Bethe and others was that it came from nuclear fusion
reactions which give off millions of times more energy than chemical
reactions. Particularly due to the work of John Bahcall, it was
realized that almost 3\% of the energy comes off in neutrinos which
could travel directly from the center of the Sun to the Earth and
provide the {\it only} direct evidence of nuclear reactions that took place in the very core of the Sun.

The pioneering experiment of Ray Davis provided evidence for solar
neutrinos but only one-third as many as the theory predicted. As a
result of a number of subsequent experiments, physicists converged
on an answer to this puzzle: two-thirds of the electron-neutrinos,
the only kind produced in the Sun, had oscillated into the other
types (muon or tau neutrinos). This conclusion was experimentally
verified by combining the results of the SNO experiment in Canada in
2000 by the direct detection of the muon and tau neutrinos in a
heavy-water Cherenkov detector, and the accurate measurement of
solar neutrinos by many other experiments. Thus the results appeared
to confirm our theory of the source of the Sun's energy and at the
same time discovered new {\it particle physics beyond the standard
model: neutrinos had mass}.

Much remains to be done in the study of solar neutrinos. Most
experiments detect the highest-energy neutrinos which are only 2 out
of 10,000 of the neutrinos that come from the sun. 85\% of solar
neutrinos come from the reaction that starts off the chain of
reactions in the sun: the weak reaction that fuses two protons
together to form deuterium with the emission of an electron and
neutrino. There is only indirect evidence for these neutrinos from
the radiochemical gallium experiments in Russia and Italy, neither
of which are able to measure the energy spectrum of these neutrinos
There are several proposals for deep underground experiments that
could detect these neutrinos and their energies. Such measurements
would provide strong direct evidence that the particular series of
fusion reactions proposed by Bethe are indeed the source of the
Sun's energy. It would indeed be wonderful if these experiments were
carried out at the Homestake site where Ray Davis pioneered the
study of solar neutrinos. A real-time measurement of pp neutrinos
can provide a 1\% measurement of neutrino mixing( $\sin ^2 \theta
_{12}$) as well as testing the sum rule connecting solar photon and
neutrino luminosities. The latter test could constrain the
possibility that subdominant neutrino sources may be present in the
Sun.

For stars larger (and hotter) than the Sun the main source of energy
is a different set of fusion reactions known as the CNO cycle. These
reactions involve carbon, nitrogen and oxygen nuclei; as a result of
their larger electric charge it takes higher energies for protons to
overcome the Coulomb repulsion and so it is most effective in hotter
stars. However calculations indicate that about 1\% of the solar
energy is produced by the CNO cycle. This produces a characteristic
neutrino spectrum at higher energies than those from the
proton-proton reaction so that proposed detectors may also be able
to detect these in spite of their low flux.  Detection of these CNO
neutrinos would be direct evidence of the CNO cycle and would also
be of great interest because it would provide information concerning
the chemical composition near the center of the Sun. Measuring the
core metallicity of the Sun could resolve some outstanding problems
with helioseismology. It would also be valuable in confirming our
theory of the CNO cycle, the major source of energy for larger
stars.

The other great event of neutrino astronomy occurred in February of
1987. In a period of 10 seconds 19 neutrinos were observed from a
type-II   supernova in two small water Cerenkov detectors: 11 in
Kamiokande in Japan and 8 in the IMB detector in the US.  These
neutrinos had been traveling for 150,000 years from outside our
galaxy and arrived a couple of years after these detectors started
to operate. From these 19 events it could be deduced that roughly 30
billion trillion trillion trillion trillion neutrinos had been
emitted in that 10 second interval. A type-II supernova is believed
to be the result of a sudden collapse of a star as a result of the
depletion of its nuclear fuel. The core of the star collapses to
nuclear density producing a huge amount of energy in the form of
highly energetic particles, but the only particle that can get out
is the weakly-interacting neutrino. Shortly after the collapse the
star begins to shine extremely bright in the sky with a luminosity
500 times that of the sun, but in fact theory tells us that 99\% of
the energy of collapse is emitted in neutrinos. This was roughly
confirmed by the 19 neutrinos of 1987.

Sometime in the future there will be a supernova explosion in our
own galaxy. This would produce many thousands of events in a
detector the size of Super-Kamiokande. As larger underground
detectors are installed they would naturally be part of a supernova
watch designed to precisely measure the properties of these many
thousands of events. Our understanding of the supernova process will
be greatly increased by observations of the energy and timing of
each of three types of neutrinos. Knowledge of the total energy
carried out by neutrinos is important to understand the proceeding
neutron star formation. The neutrinos we observe will clearly be
affected by oscillations so that it will be better if we have good
information on oscillation parameters. There may be a large
sensitivity to $\sin ^2 \theta _{13}$, the value of which may be
determined by reactor neutrino experiments. We need to be able to
make precision measurements of cooling time, timing of
neutronization pulse, average energy and time-integrated luminosity
and the neutrino spectrum. It is worth noting that neutrino
observations of core-collapse supernovae and merger of neutron stars
and other compact objects also provide a probe complementary to
other U.S. scientific investments such as LIGO.

It is believed that in a core-collapse supernova many of the heavy
elements in nature are first formed by what is called the r-process.
The abundance of different elements and isotopes depends on the
neutron-proton ratio at the site and this is influenced by the
neutrino flux. Thus observations of supernova neutrinos may help us
better understand the origin of elements. Megaton-scale detectors
with a reach of about 10 megaparsecs may enable to search for
supernovae in other galaxies without necessarily seeing the
accompanying photons.  While a supernova in our galaxy may not occur
for many years it will be possible in megaton detectors to see a
small number of neutrinos from many other galaxies. Thus interesting
statistics on supernova neutrinos can be acquired.  Core-collapse
supernova physics nicely illustrates that astrophysical extremes
allow testing neutrino properties in ways that cannot be done
elsewhere, e.g. exploring the neutrino-neutrino interaction effect
as an ``emergent phenomenon".

Supernovae have been exploding for billions of years and they supply
a diffuse background of neutrinos. The detection of the higher
energy neutrinos in this background should be possible in a large
deep underground detector. This will allow us to look back in time
and gain information on the rate of star formation in the distant
past, a unique window on the history of our Universe.

This is a new kind of astrophysics looking at the interior of
compact objects. Even though neutrino observations are typically
complementary to observing light, this does not necessarily have to
be the case: There could always be something new. Finally it should
be pointed out that neutrino astrophysics is one of the most
interesting basic science topics suitable for terascale computing
applications

\newpage

\section{Dark Matter Detector}

\subsection{Executive Summary}

The Universe is filled with a mysterious form of dark matter, the
understanding of whose fundamental nature poses the greatest
challenge to contemporary cosmology and particle physics. We argue
that the best motivated dark matter models are on the verge of being
explored by direct dark matter search experiments that would
enormously benefit from a laboratory like DUSEL.  Dark matter
searches in DUSEL offer a unique opportunity for ground-breaking New
Physics discoveries. The future laboratory will play a fundamental
role in post-discovery dark matter studies, and in giving the United
States a leading role in one of the hottest field in physics today.

\subsection{Dark Matter}

The nature and identity of matter in the Universe is  one of the
most persistent questions ever posed and one of the most challenging
problems facing modern cosmology and particle physics. The wealth of
evidence for a new form of matter from many very different types of
observations is overwhelming. The first indication for dark matter
came from the observations of high velocities of galaxies within
clusters of galaxies which required an additional source of gravity
beyond that which could be accounted for from light-producing
galaxies in the cluster.  Similarly, the rotation of stars and gas
in spiral galaxies also pointed to the notion that galaxies were
embedded in a large and massive halo of dark gravitating matter.

More recent observations have bolstered the necessity for the
existence of dark matter.  Large amounts of gas (mostly hydrogen and
helium) around galaxies and clusters of galaxies tend to be very hot
(of order a few million degrees). At these temperatures, the gas
emits electromagnetic radiation in the X-ray frequency bands which
have been observed by X-ray satellites. The presence of this gas
requires a significant amount of gravity to prevent the hot gas from
flying off into space, in turn yet another indirect evidence for
dark matter.

It is also known that gravity can bend the path of light.
Observations of distant galaxies along the line of sight of a large
cluster of galaxies shows clear signs of the gravitational lensing
of the light.  Indeed, recent claims to direct evidence for dark
matter came from observations of a collision of clusters of galaxies
showing lensing which is directly associated with (non-dissipative)
dark matter in tact with the two clusters whereas the (dissipative)
hot gas has been stripped from the cluster by the collision.

Our theoretical understanding of the very existence of galaxies and
structure in the Universe itself also relies on the existence of
dark matter.  The initial seeds or perturbations in an otherwise
smooth and homogeneous Universe could not have grown sufficiently to
produce galaxies  and clusters without a dark matter component.
Evidence for the imprint of these seeds have been left on the
microwave background radiation, and detailed measurements by many
experiments have precisely determined several of the key parameters
describing our Universe including the abundance of dark matter. All
of these measurements are completely consistent both qualitatively
and quantitatively with the fact that dark matter dominates the
overall matter budget of the Universe.

All of the experiments and astronomical observations which have
established the existence of dark matter are to date incapable of
determining the identity and fundamental nature of the dark matter
particle. In principle, one could imagine that the dark matter is
simply some form of ordinary matter ({\em i.e.} made of neutrons and
protons), in a non-luminous state (such as a dead star or perhaps
dust).  While there are many astrophysical constraints against such
objects, our understanding of the early Universe through
nucleosynthesis (the process of formation of the light elements,
Deuterium, Helium, and Lithium) places a firm limit on the total
amount of normal matter which is far below the requisite amount of
dark matter. Furthermore detailed analysis by the WMAP experiment
observing microwave background anisotropies has firmly established
the relative amounts of total matter (normal plus dark) to normal
matter.

Another possibility for dark matter among known particles in physics
is the neutrino, now known to have a small but finite mass.
However, it has also been established that the neutrino abundance
can comprise less than 1\% of the matter density.

\begin{figure}[ht]
\includegraphics[width=6in]{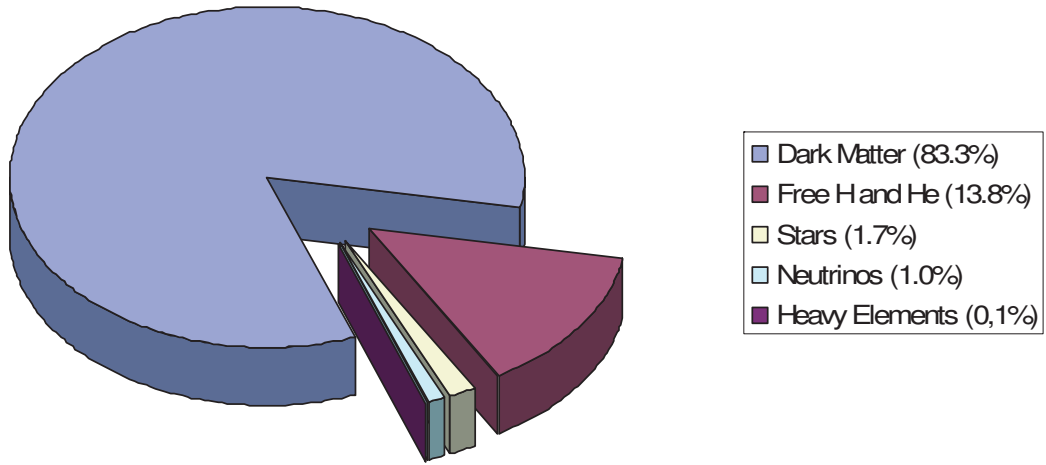} \caption{The pie chart shows
the relative abundance of different components of matter.}
\label{fig:pie}
\end{figure}

We are led to the conclusion, therefore, that the predominant form
of matter in the Universe can not be composed of normal matter ({\em
i.e.} made of protons and neutrons) and must be related to physics
beyond the standard (and well established) model of strong and
electro-weak interactions. The distribution of the known components
of matter with the abundance of dark matter ({\em i.e.} excluding
dark energy, the non-clustering component of the Universe)  is shown
in Fig. \ref{fig:pie}.

The theoretical literature contains many new physics models that
also contain candidate cold dark matter (CDM) particles. Some of the
many candidate CDM particles include: axions, sterile neutrinos,
weakly interacting massive particles (WIMPs), SuperWIMPs, $Q$-balls,
black hole remnants and fuzzy cold dark matter. Out of this list,
axions and WIMPs stand out in that they occur as by-products of
theories which solve longstanding theoretical problems with the
standard model. Axions occur in models which solving an important
quantum problem associated with the strong nuclear force.  Axions
can be searched for in terrestrial experiments using microwave
cavities. WIMP particles typically occur in theories which attempt
to explain the mechanism behind the breakdown of the electro-weak
symmetry (the symmetry relating electromagnetism to the weak nuclear
force). WIMPs are especially compelling in that a calculation of
their relic abundance from production during the Big Bang falls very
close to the measured abundance if the particles are weakly
interacting, and have mass of order 100 times the proton mass and is
of order the weak scale. This fact can be construed as independent
{\it astrophysical} evidence for the existence of new physics at the
weak scale.

There are many examples of WIMP particles. Candidates arising from
supersymmetric (SUSY) theories-- due to their deep connection with
grand unified and superstring theories, along with some indirect
experimental support-- have been most thoroughly examined. SUSY
theories posit that each known particle has a superpartner with mass
of order the weak scale. The superpartners arise due to the
intrinsic Fermi-Bose symmetry that is the foundation of SUSY
theories.

The CERN LHC which just turned on---with proton-proton
collisions at a record 14 trillion electron Volts-- should have
enough energy to produce the superpartners directly in a laboratory
environment. The superpartners are expected to decay via a cascade
of particles which includes the lightest SUSY particle, which is the
CDM candidate. The presence of collider events with large missing
energy beyond that expected from the standard model would signal the
production of dark matter at the LHC. A measurement of various
properties of the superparticles would allow one to determine the
nature of the SUSY model, and measure properties such as the mass of
the CDM particle.

Relic WIMPs can also be detected indirectly. In one instance, the
sun, as it follows its path through the galaxy, may actually sweep
up WIMP particles which then collect at high density in the solar
core. The high density core WIMPs may then annihilate one with
another into standard model particles, including neutrinos with very
high energies. The proton decay search facility to be housed in
DUSEL will also have dramatic consequences for dark matter searches,
as it would allow to look for these energetic neutrinos produced in
WIMP annihilations in the center of the Sun or of the Earth. In many
respects, this facility would go beyond the reach of experiments
like IceCube, situated at the south pole, in this search technique,
and be complimentary to direct detection.

WIMPs circulating in the galactic halo may also occasionally
annihilate one-with-another. Annihilation products such as gamma
rays and anti-matter particles can be detected at various ground and
space-based detectors. Gamma rays have the advantage in that they
would point directly back to the spot at which the annihilations
occur. A knowledge of WIMP properties gained from direct detection
and collider searches, combined with gamma ray data from WIMP halo
annihilations could allow for dark matter tomography measurements of
the galactic dark matter density distribution.

Dark matter particles left over from the Big Bang may also be
detected directly via their collisions with nuclei in low background
environments located deep underground (generally, the deeper the
better), an outstanding example being DUSEL. In fact, direct WIMP
detection is in every way complementary to detection at the LH---in part because direct detection would verify that the missing
energy particles found in LHC collisions would actually be dark
matter particles---and to indirect detection, being free of the
complicated intrinsic astrophysical backgrounds that hinder the
possibility of firmly establishing an anomalous indirect signal as
coming from dark matter.

The field of direct dark matter detection has recently succeeded in
achieving major progress in sensitivity and in demonstrating the
feasibility to scale up the size and performance of current
experiments. In addition, several different techniques are being
successfully explored, including solid state, noble gas and bubble
chamber detectors. The CDMS-II and Xenon-10 collaborations --
employing two different target nuclei---reported comparable
extremely competitive new sensitivity limits. These new limits are
triggering a lively reaction in the particle and astro-particle
theory community, as for the first time well-motivated particle dark
matter models are being explored and constrained.

While it is hard to quote a model-independent {\em lower limit} on
the expected signal from scattering of dark matter off nuclei, the
exploration of the parameter space of theoretically motivated
particle dark matter theories and independent arguments based on the
observed abundance of dark matter in the Universe indicate that a
signal is to be expected in the range between the current
sensitivity and what is anticipated to be possible in experiments at
DUSEL. The generic expectation for some of the best understood and
motivated particle physics models for CDM---including SUSY---is
that a large fraction of models lies between the current sensitivity
and the sensitivity that current experimental techniques could
achieve in DUSEL at the 4850 ft depth level. Most of the models with
a faint predicted signal should be explored in DUSEL at the 7400 ft
depth level with large (ton-sized) detectors. Even a non-detection
would have profound consequences if such sensitivities are achieved.

The most general possible interaction between a nucleus and a WIMP
includes in the non-relativistic limit a coherent spin-independent
coupliing (scaling as $A^2$, where $A$ is the target nucleus atomic
number) plus a spin-dependent coupling. In general, theory indicates
that spin-independent scattering on large $A$ nuclei is the most
promising direct dark matter detection technique. Models exist,
however, where the spin dependent cross section is a better
detection channel.

The implications of direct detection of dark matter would be
formidable, and would open a new chapter in the history of Science.
The high-energy and astro-particle theory community believes that
DUSEL will be important---if not fundamental---both at the stage
of discovery and in the subsequent phase of study of the properties
of the dark matter particle. If a dark matter signal is detected,
the discovery will need to be both confirmed by other experiments
and reproduced. As the reported detection of dark matter by the DAMA
collaboration has taught the community, systematic uncertainties and
backgrounds are a problematic issue, and likely depend on
environment: in this respect it will be essential to have more than
one facility where to run direct dark matter detection experiments.
Having a large underground laboratory will allow a multi-pronged
approach, which will make use of several experimental techniques and
different target materials. This will entail the possibility of
confirming a detected signal rate, if studying the recoil spectral
shape, investigating the $A^2$ scaling with nuclear number. In turn,
This will ultimately make it possible to measure the interactions of
dark matter with ordinary matter and to estimate the particle dark
matter mass independently of colliders. In addition, any claim of
periodicity (day/night or seasonal) will benefit from the
simultaneous operation of two or more detectors.

As alluded to above, models exist where spin-dependent searches are
the only way to detect a dark matter signal. Having at disposal a
large underground facility would allow to set up this second
category of experiments, whose performance has been shown to be in
many ways complementary to spin-independent searches, and whose
results can yield important information on the spin of the dark
matter particle. Furthermore, it was shown that measuring the ratio
of spin-independent to spin-dependent scattering rates with at least
two targets can lead to the discrimination of, for example,
supersymmetric and extra-dimensional models.

Along similar lines, a deep underground laboratory might allow the
development of dark matter search experiments requiring strong
cosmic ray background suppression that could explore alternative
dark matter particle scenarios. For instance, special experiments
would be needed to explore super-light dark matter scenarios, and if
underground space allows, could be hosted in DUSEL.

The detection of a signal will usher in an era of dark matter
astronomy, in which we will be able to study the dark sector of the
Universe directly. It will become feasible to directly observe the
structure of the dark matter halo, and answer questions like: how
much dark matter is there in our galaxy, in particular near the
Solar System? Is dark matter concentrated in clouds or is it
smoothly spread out all over the galaxy? Are there streams of dark
matter like our current theories predict? Data from direct detection
experiments at DUSEL will be combined with other data on the
galactic halo to arrive at an understanding of how our galaxy
formed. For example, the GLAST space-born gamma-ray telescope may
detect the emission from dark matter in the halo, and observational
programs in the Sloan Digital Sky Survey will detect and measure
stellar structures and motions in the galactic halo.

In a possible scenario, the Large Hadron Collider discovers a new
particle that theorists determine is suitable to be the dark matter.
At the same time, direct detection experiments at DUSEL observe
particles from outside the Solar System and confirm that the
particles discovered at the Collider is indeed the dark matter.
While dark matter experiments at DUSEL go on to measure the density
of dark matter in the vicinity of the Sun, other indirect dark
matter searches map the amount and location of dark matter in our
galaxy. Further directional detectors at DUSEL measure the
velocities and arrival directions of the dark matter particles.
Theoretical studies ascertain that these velocities are consistent
with the results of indirect searches. This brings direct evidence
of streams of dark matter swirling in the galactic halo, directly
confirming the idea that galaxies are formed by combining small dark
matter clouds into larger and larger structures.

The theory community is greatly excited by the prospective of DUSEL
since this would create a center for dark matter studies in the
United States. Laboratories have always, in the history of particle
physics, triggered very fruitful interactions among theorists and
between theorists and experimentalists. Theory workshops could be
hosted at DUSEL, as well as yearly executive summaries of
theoretical progress for experimentalists and of experimental
progress for theorists.

Finally, in the forthcoming era of large direct detection
collaborations, it will be strategic for the US scientific community
to have a laboratory like DUSEL in the US, preventing the risk of a
drain of human resources outside the country, with dangerous
consequences in particular for the theory community.

\newpage

\section{Neutrinoless Double-Beta Decay}
\subsection{Executive Summary}

The neutrino is an elementary particle that scatters only through
the weak interaction, and consequently rarely interacts in matter.
Neutrinos carry no electric charge, have spin one-half, and belong
to a family of particles called leptons. Leptons are particles
which interact via weak, electromagnetic or gravitational forces
only. The electron and electron neutrino are leptons with lepton
charge, $L = +1$.  Their anti-particles have lepton charge, $L =
-1$. Neutrinos differ from quarks, which have both strong and
electromagnetic interactions, and from other leptons, such as the
electron, which are charged and thus interact electromagnetically.

Owing to the rarity of their interactions, the neutrinos are
elusive. Studying them experimentally is a major challenge. While
their existence was first postulated almost 80 years ago, it took 25
years to detect them in the laboratory.  Only recently have
high-precision neutrino experiments been possible.  One fact
established from experiments of the past decade is that, contrary to
the prediction of the otherwise well confirmed standard model of the
elementary particles, neutrinos are not massless. But their masses
are extremely tiny: a neutrino is at least a million times lighter
than the next lightest particle, the electron.

The most natural explanation for the lightness of neutrinos,
compared to all other particles,  is the so-called See-Saw
Mechanism.  The See-Saw Mechanism is only possible for electrically
neutral particles, which can, in principle, be their own
anti-particles.  The mass term for such a particle can include what is known as a
Majorana mass.  The irony of the See-Saw Mechanism is that neutrinos are so
very light due to the existence of a very heavy mass, $M \gg M_W$.
In fact, $M$ is so very large that it is close to the grand
unification scale where the electromagnetic, weak and strong
interactions are equal. In fact neutrino masses may be another
indication of grand unification and supersymmtry. Neutrinoless
double beta decay experiments can prove that the neutrino has a Majorana mass and is its own anti-particle.

\subsection{Neutrinoless Double Beta Decay and the Origin of Mass}

The question of the origin of masses, for the neutrinos and
everything else, is one of the central puzzles of elementary
particle physics. Several observations---the extreme lightness of
the neutrinos, their electrical neutrality, and important
theoretical principles---suggest that the mechanism generating
neutrino masses differs from that generating other particle masses.
Because neutrinos are neutral, a neutrino can serve as its own
antiparticle: consequently, unlike other leptons and unlike the
quarks, neutrinos can possess so-called Majorana masses.  This kind
of mass, when added to the Dirac mass that neutrinos (and other
Standard-Model particles) can have, can account (through the See-Saw Mechanism) for the tiny but nonzero neutrino masses found experimentally.

The only known feasible way to try to confirm that the neutrino has
a Majorana mass (and thus lacks a distinct antiparticle) is to
search for neutrinoless double beta decay---the process $(N,Z)
\rightarrow (N-2,Z+2) + e^- + e^-$, in which a parent atomic nucleus
with $N$ neutrons and $Z$ protons decays to a daughter nucleus
$(N-2,Z+2)$ plus two electrons. The observation of this process at
any nonzero level would establish that neutrinos do possess Majorana
masses and that they are indeed their own antiparticles. Majorana
neutrino masses, if present, must arise from new physics beyond the
standard model, very likely at a high mass scale far beyond the
scale described by the standard model, and far beyond that
accessible to the Large Hadron Collider (LHC) at CERN. Thus,
neutrinoless double beta decay seeks mass-related new physics that
would be invisible to the LHC.

One of the most conspicuous features of the Universe is that it
contains atoms, of which we are made, but essentially no anti-atoms,
which, had they been present, would have annihilated us. A leading
candidate for the explanation of this crucial asymmetry is
Leptogenesis, which states that the asymmetry arose from the decays
of very heavy neutrinos that existed in the early Universe.
Leptogenesis is an outgrowth of the See-Saw Mechanism, the leading
candidate for the explanation of the incredible lightness of today's
neutrinos. A signature feature of the See-Saw Mechanism and
Leptogenesis is the prediction that neutrinos are their own
antiparticles. Thus, the observation of neutrinoless double beta
decay would provide evidence supporting both the See-Saw Mechanism
and Leptogenesis.

The observation of neutrinoless double beta decay, $(N,Z)
\rightarrow (N-2,Z+2) + e^- + e^-$, would also demonstrate that {\em
lepton charge} is not conserved : a final state containing two
leptons is produced from an initial state containing none. This
non-conservation would remove one of the principles protecting the
proton from decaying into lighter particles. Slow proton decay,
predicted by the theories that seek to unify the forces of nature,
would eventually leave the Universe a very different place than it
is now. Confirmation that protons do decay is a major goal of the
very large underground detectors that would also study neutrinos
from a distant accelerator.

It is likely that neutrinoless double beta decay, if it occurs, is
dominated by physics that makes its rate proportional to $m_{\beta
\beta}^2$, where $m_{\beta \beta}$, the effective Majorana mass for
neutrinoless double beta decay, is a linear combination of neutrino
masses. Thus, a measured value of the rate would provide information
on neutrino masses---information of a kind that cannot come from
experiments on neutrino oscillations. While all of our current
evidence for neutrino mass comes from oscillations, such experiments
can only provide information on mass squared differences, not on
absolute masses.

\begin{figure}
\includegraphics[width=5in]{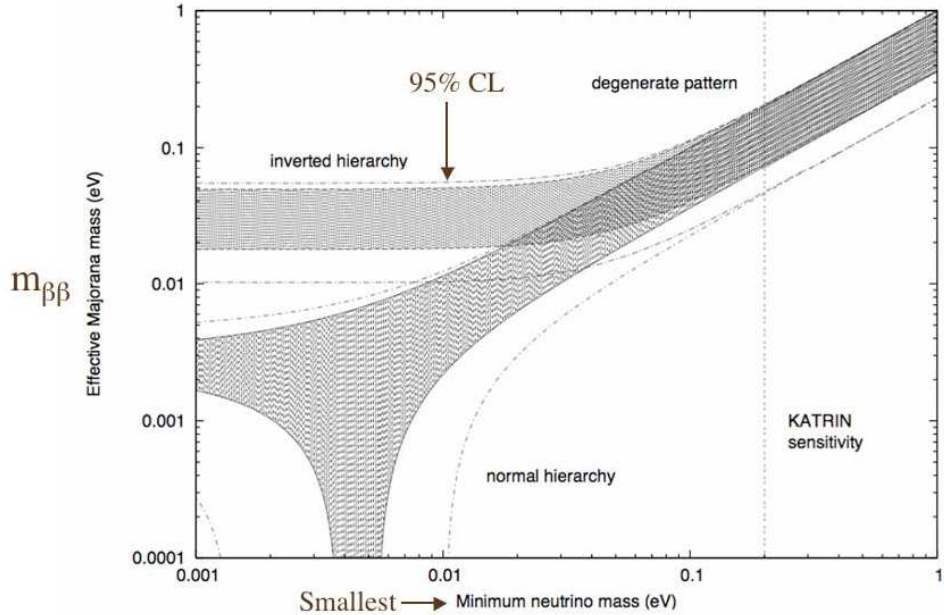}
\caption{The effective neutrino mass $m_{\beta \beta}$ observable in
neutrinoless double beta decay experiments as a function of the
smallest neutrino mass for both the normal and inverted neutrino
mass hierarchy.} {\label{fig:0nubeta}}
\end{figure}

The sensitivity that an experiment needs to achieve in order to
detect neutrinoless double beta decay depends on the value of
$m_{\beta \beta}$. The possible values of $m_{\beta \beta}$, in
electron volts, are shown in Fig. \ref{fig:0nubeta} as a function of
the mass of the lightest neutrino. This figure assumes that there
are three distinct neutrinos of definite mass, an assumption in
accord with experiments done to date.  From neutrino oscillation
data, we know that the mass squared difference between the first and
second neutrino is about 30 times smaller than that between this
pair and the third neutrino. These results allow two mass patterns.
If the closely spaced pair is at the bottom of the spectrum, the
neutrino mass pattern is called the {\em normal} hierarchy, as it
then resembles the quark spectrum.  The alternative---the closely
spaced pair at the top---is called the {\em inverted} hierarchy. If
all neutrinos were much heavier than their mass differences, both
the normal and inverted hierarchies would appear as three nearly
degenerate neutrinos.

>From Fig. \ref{fig:0nubeta}, we see that if the hierarchy is
inverted, $m_{\beta \beta}$  cannot be smaller than 0.01 eV. From
the standpoint of theory, the inverted and normal hierarchies are
equally likely. Thus, a very reasonable goal for the next generation
of experiments is to be sensitive to $m_{\beta \beta}$ down to 0.01
eV, so that the inverted-hierarchy possibility will be fully
covered.

The rate for neutrinoless double beta decay of a particular parent
nucleus is proportional, not only to $m_{\beta \beta}^2$, but also
to the square of a nuclear matrix element involving the wave
functions of the parent and daughter nuclei and the nuclear operator
for double beta decay. This matrix element must be taken from
theory, which presently is uncertain by about a factor of two,
judging from the spread in various predictions. Consequently, there
is a significant theory uncertainty in estimating the decay rate
that would correspond to a given value of  $m_{\beta \beta}$.
Neutrinoless double beta decay rates also vary substantially from
nucleus to nucleus due to differences in the energy released in the
decay.  But in favorable cases where the energy release is large,
one expects neutrinoless double beta decay lifetimes on the order of
$10^{28}$ years for $m_{\beta \beta} \approx 0.01$ eV. This long
lifetime, longer than any decay lifetime measured in any process to
date, makes searches for neutrinoless double beta decay very
challenging. Experiments require very large detectors, e.g., 1-10
tons of the parent isotope, and extraordinary efforts to eliminate
backgrounds from cosmic rays and natural radioactivity. Achieving sensitivity to much
of the $m_{\beta \beta}$ region allowed by a {\em normal} hierarchy
will require still larger experiments, ones with hundreds of tons of
parent material.

If the hierarchy is known to be inverted, and one achieves sensitivity to 0.01 eV but finds no double beta decay, then either neutrinos are not their own antiparticles, or some interesting new physics beyond that which leads to Fig. \ref{fig:0nubeta} is at work.
Neutrinoless double beta decay is sensitive not only to neutrino
mass, but also to various exotic forms of lepton charge
violation that are found in some extensions of the standard model.
Several of these exotic mechanisms are connected with
supersymmetry, a symmetry predicting the existence of many new
particles, partners of those found in the standard model.  Finding
these new particles is one of the major goals of the LHC.  Were the
LHC to make such a discovery, there would be intense interest in
neutrinoless double beta decay mechanisms that arise from supersymmetry.

Experiments currently under development hope to cover most of the
$m_{\beta \beta}$ region allowed by the inverted hierarchy.  Both
the technologies employed and the parent isotopes chosen for these
experiments are quite varied.  The latter is important, eliminating
the possibility that an unfortunate parent isotope choice, one with
a suppressed nuclear matrix element, could invalidate the
conclusions of an experiment.  The best current limits on
neutrinoless double beta decay come from experiments using germanium
detectors enriched in the double-beta decay isotope of interest,
$^{76}$Ge.   New experiments scaling up this technology by factors
of 10 to 100 are under development in both Europe and the US. Other
new efforts include cryogenic arrays of ultra-sensitive TeO$_2$
bolometers, in which the decay of the parent isotope, $^{130}$Te,
can be seen, and a liquid xenon detector whose prototype is under construction at WIPP.
The xenon detector, when fully developed, will employ a novel atomic
physics technique to detect the daughter nucleus $^{136}$Ba produced
in the decay of $^{136}$Xe, making this experiment unusually
insensitive to backgrounds.  While many other efforts are underway,
the three mentioned here include many U.S. physicists among the
proponents.

If such next-generation experiments fail to detect neutrinoless
double beta decay, next-to-next generation experiments at the 100-ton scale would likely become the next goal.  Indeed, some of the
experiments now under development hope to be scalable to still
larger masses.  Alternatively, if next-generation experiments
discover neutrinoless double beta decay, more demanding follow-up
experiments would still be needed, to probe the underlying
mechanism.  More information can be gained by studying which
daughter-nucleus states are populated in the decay and by measuring
the energy spectrum and angular correlation of the outgoing
electrons.

Because next-generation experiments will be so challenging, it is
important to have multiple efforts, to guard against an unidentified
background producing a false result.  Were several experiments to
find neutrinoless double beta decay, this would not only confirm the
result but also help theorists translate the rates into a reliable
$m_{\beta \beta}$. Uncertainties in estimated nuclear matrix
elements can be better assessed if predictions connected with a
single $m_{\beta \beta}$ can be tested in several nuclei.

In summary, the search for neutrinoless double beta decay probes the
physics of mass, in a way that is completely different from, and
complementary to, the LHC program. A positive result would establish
that neutrinos have Majorana masses and are their own antiparticles,
distinguishing them from all the other constituents of matter. It
would show that lepton charge is not conserved.  The
observation of neutrinoless double beta decay would demonstrate the
physics necessary to both the See-Saw Mechanism and Leptogenesis. It
would also remove one of the protections against proton decay. The
search for neutrinoless double beta decay is inherently
interdisciplinary, involving techniques and concepts from particle,
nuclear, and condensed matter physics. To date, apart from one
controversial claim, neutrinoless double beta decay has not been
observed.  But we now know that an interesting mass range is within
reach of next-generation detectors, given what has been learned
about neutrino mass from oscillation experiments. These detectors
will not only be of unprecedented scale, but also require a degree
of background suppression not yet achieved in physics.

\newpage

\section{Neutron-Anti-Neutron Oscillations}

\subsection{Executive Summary}

The process of neutron--antineutron ($n - \bar n$) oscillation is an
important way to probe the basic instability of matter that is
believed to be at the heart of our understanding the origin of
matter in the Universe as well as the nature of new forces
responsible for neutrino masses. It complements the searches for
proton decay, which have been conducted for the past two decades and
are ongoing. At the Deep Underground Science and Engineering
Laboratory, $n - \bar n$ oscillation time scales of order $10^{10}$
sec. can be probed, with the potential to reveal answers to many
fundamental questions in elementary particle physics, with
implications for nuclear physics, cosmology and astrophysics: Is the
neutron its own antiparticle? What is the degree of instability of
matter? What is the basic mechanism for the creation of matter over
antimatter in the Universe?  Are there extra space--like dimensions?
Is there a hidden ``parallel" Universe? What is the nature of dark
matter in the Universe?

\subsection{Introduction}

A key concept that has emerged from recent theoretical studies
seeking the ultimate unity of matter and forces is that at the
fundamental level, matter is predicted to be unstable. The apparent
stability of the Universe is a consequence of the fact that matter
instability occurs on a time scale which is more than a trillion
trillion ($10^{24}$) times the age of the Universe. A further
argument that reinforces this belief is the realization during the
past four decades that matter instability is indeed essential if one
wants to understand why the observed Universe is made only of matter
and no antimatter. The challenge for physics is to discover how
matter instability manifests itself and what the degree of this
instability is.

\begin{figure}[h]
\begin{center}
\includegraphics[width=0.7\linewidth]{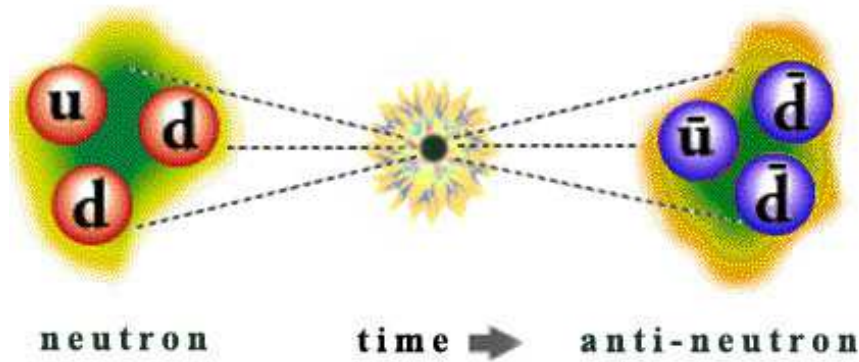}
\vspace*{-0.2in}\caption{Neutron oscillating into anti-neutron.}
\label{fig:nnbar2}
\end{center}
\end{figure}

There are two known ways in which matter instability can be
manifest: (i) the decay of a proton (or a neutron which is otherwise
stably bound in a nucleus), discussed in another part of the white
paper, and (ii) spontaneous conversion of neutron ($n$) to
anti--neutron ($\bar{n}$), called $n-\bar{n}$ oscillation (see Fig.
\ref{fig:nnbar2}), the subject of this part of the white paper.
Spontaneous conversion of other electrically neutral particles such
as K-,B-meson into their antiparticles has already been
experimentally established, providing ground-breaking information
about the fundamental forces and constituents of matter. They have
guided the course of elementary particle physics for the past
half-century. The $n-\bar{n}$ oscillation is even more profound and
is expected to provide insight into many fundamental issues
confronting particle physics today.

It was pointed out in the early 1970's that since a neutron ($n$) is
electrically neutral, it could convert itself to an anti-neutron
($\bar{n}$), and, moreover, that this conversion process could
provide a way to understand the observed fact that there is an
asymmetry between the amount of matter and antimatter in the
Universe. In the early 1980's, reasonable and consistent particle
physics models were discovered which predicted that neutrons are
their own anti-particles, similar to massive neutrinos being their
own anti-particles, and that $n-\bar{n}$ oscillation should occur at
an observable rate. This led to increased experimental as well as
theoretical interest in this process. The existence of free
$n-\bar{n}$ oscillations would also mean that neutrons inside nuclei
would become anti-neutrons and make nuclei unstable. However, due to
the difference between the way neutrons and anti-neutrons behave in
the presence of nuclear forces, matter stability is highly
suppressed. As a result, oscillation times of a year would
correspond to about $10^{30}$ years for nuclear instability.
Ongoing proton decay experiments also found lower limits on nuclear
instability time scales due to $n-\bar{n}$ oscillation in nuclei in
the same range. It was realized that available reactor neutrons
could be used to probe $n-\bar{n}$ oscillations with oscillation
times precisely in the range of a year to few years. There are,
however, severe limitations to discovering $n-\bar{n}$ oscillations
inside nuclei, due to atmospheric neutrino backgrounds.
Uncertainties concerning the relevant nuclear properties also make
it difficult to pin down precisely the value of the oscillation
time.  Thus it seems that the most promising way to search for this
process further is to search for $n - \bar n$ oscillations with free
neutrons. This is what we propose to carry out at the DUSEL
facility.

\subsection{Present experimental situation}

The only free $n-\bar{n}$ oscillation experiment carried out to date
was at the European laboratory at Institut Laue-Langevin (ILL),
where a lower limit on the oscillation time scales of $\tau_{n-\bar
n} > 0.86 \times 10^8$ sec (90 \% CL) was established.  This lower
limit can be interpreted as follows.  A free neutron, which itself
is unstable with a mean lifetime of $\tau_n = 886 \pm 1$ sec., does
not oscillate to an anti-neutron in the time span of $0.86 \times
10^8$ sec. (or 2.7 years). There are theoretical reasons to believe
that an oscillation time which is a factor of 100 above this limit
could probe some very interesting new ideas in physics beyond the
standard model.

As noted earlier, the oscillation rate of a free neutron in vacuum
can be much faster than that inside a nucleus, where it is
suppressed by internuclear forces. In fact, analysis of $n-\bar n$
oscillations in matter yields the relation between the free neutron
oscillation time and the nuclear neutron oscillation time given by
$\tau_{nuc} = R \tau^2_{n-\bar{n}}$, where $R$ is a
nucleus-dependent factor. Detailed nuclear physics calculations
yield $R({}^{16}{\rm O}) \simeq 0.5 \times 10^{23}$ s$^{-1}$ and
$R({}^{56}{\rm Fe})\simeq 0.7 \times 10^{23}$ s$^{-1}$. So the
current lower bound on the free neutron oscillation time,
$\tau_{n-\bar n}
> 0.86 \times 10^8$ sec., corresponds to a lower limit on the nuclear decay
time of about $\tau_{nuc} \gtrsim 1 \times 10^{31}$ yrs. Recent and
ongoing proton decay searches are sensitive to similar lifetimes for
the $n-\bar{n}$ oscillation mode.  The Soudan-2 experiment has
reported a lower limit $\tau_{nuc} > 0.72 \times 10^{32}$ yrs.,
which corresponds to the lower bound $\tau_{n-\bar n} \gtrsim 2
\times 10^8$ sec. Therefore, such an apparently small free neutron
oscillation time is not in conflict with matter stability bounds.

\subsection{Why is it important to conduct a high sensitivity search for
$n-\bar{n}$ oscillation at DUSEL?}

There are several reasons to conduct a high-sensitivity search for
$n-\bar{n}$ oscillations, as enumerated below. DUSEL would serve as
an ideal site for this search.

\vspace*{0.15in} \noindent {\bf Why are the conventional proton
decay search experiments not adequate?} \vspace*{0.15in}

\begin{figure}[h]
\begin{center}
\includegraphics[width=0.7\linewidth]{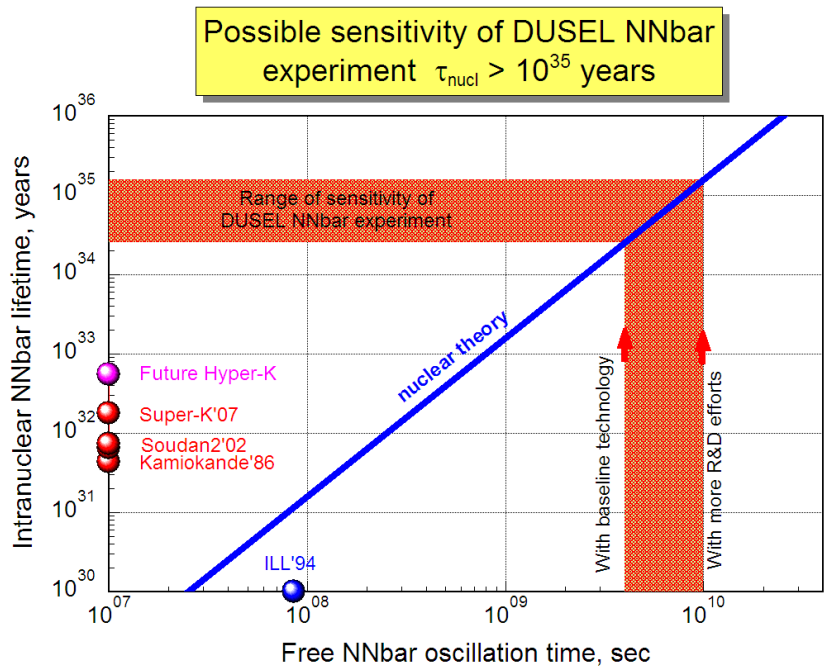}
\caption{Sensitivity of DUSEL NNbar experiment on matter instability
lifetime.} \label{fig:nnbar1}
\end{center}
\end{figure}

If $n-\bar{n}$ oscillations exist, a neutron inside the nucleus
could oscillate to an anti-neutron, which would subsequently
annihilate with surrounding nucleons to give typically five pions
with an invariant mass of two GeV. This will make nuclei unstable.
Thus, in principle, conventional searches for proton decay (and
decays of neutrons otherwise stably bound in nuclei) can probe for
$n-\bar{n}$ oscillations. The relation $\tau_{nuc.} = R
\tau^2_{n-\bar{n}}$ enables one to deduce the limit or value for
$\tau_{n-\bar{n}}$ from the limit or value for the nuclear
instability time $\tau_{nuc}$. However, proton decay experiments are
not a very sensitive way to probe $\tau_{n-\bar{n}}$, compared to a
direct search, for the following reasons. The first point is that an
order-of-magnitude increase in $\tau_{nuc}$ only leads to a factor
of three improvement in $\tau_{n-\bar{n}}$. More importantly,
because of the presence of backgrounds, the lower limits on
$\tau_{nuc}$ that can be derived go at most like the square root of
the exposure time.  Hence, if the fiducial volume of the proton
decay search detector increases by a factor of 25 (as is being
contemplated for the next generation of proton decay searches), the
lower limit on $\tau_{nuc}$ will increase at most by a factor of
five, and the limit on $\tau_{n-\bar{n}}$ will go up only by a
factor of 2.4.  The published Soudan-2 limit on $\tau_{nuc}$ yields
$\tau_{n-\bar{n}}\gtrsim 2\times 10^8$ sec., and the preliminary
Super-K limit yields $\tau_{n-\bar{n}}\gtrsim 3\times 10^8$ sec. The
$\tau_{n-\bar{n}}$ reach of the planned experiments to search for
proton decay is only about $\tau_{n-\bar n} \simeq 7\times 10^8$
sec, which is far lower than the reach of $\tau_{n-\bar n} \simeq
10^{10}-10^{11}$ sec anticipated for the free neutron search
experiments being contemplated. Fig. \ref{fig:nnbar1} displays the
current and future sensitivity of matter instability lifetime from
$n-\bar{n}$ oscillations as well as from proton decay searches.

\vspace*{0.15in} \noindent {\bf What can we learn about the
fundamental forces in nature and the working of the Universe from
the ${n-\bar{n}}$ search?}\vspace*{0.15in}

Neutron-anti-neutron oscillations touch upon many areas of physics
ranging from elementary particle physics to nuclear physics to
astrophysics and cosmology. The potential of the search for $n -
\bar n$ oscillations as a probe of physics beyond the standard model
is comparable to that of neutrino oscillations, which led to the
very important discovery of neutrino masses and mixing. We now
enumerate the most significant questions that will be addressed by
an $n-\bar{n}$ oscillation search.

\begin{enumerate} \item {\bf  Is the
neutron its own anti-particle?} \vspace*{0.15in}

A search for $n-\bar n$ oscillations will probe new forces among
particles at distance scales about a hundred times shorter than the
ones that will be probed by the Large Hadron Collider (LHC).  In
specific models with low energy supersymmetry, the length scale of
new physics probed in an $n-\bar{n}$ oscillation search can be one
hundred million times shorter than the ones probed by the LHC.  The
discovery of neutrino masses is already providing one such probe of
extremely tiny distances.  Neutrinoless double beta decay
experiments aim to test whether the neutrino is its own
anti-particle, which is implied by many appealing theories
explaining its small mass.
If this is the case, then there are reasons to think that the neutron is
its own anti--particle,
a suggestion which goes back to the classic paper of
Ettore Majorana in 1930's, since the neutrino and the neutron are linked
by a symmetry of the standard model called B-L symmetry (see below)
 This gives
greater motivation for the existence, at an observable level, of the
phenomenon of $n-\bar{n}$ oscillation which is the experimental
manifestation of the neutron being its own anti-particle.

\item {\bf  A new fundamental symmetry
probed by $n-\bar{n}$ oscillations} \vspace*{0.15in}

To see which symmetry is probed in $n-\bar{n}$ oscillation searches,
it is useful to compare with the situation involving neutrinos
again. If the neutrino is its own antiparticle, this breaks total
lepton number, $L$, by two units ($|\Delta L|=2$).  At an earlier
time, most physicists believed that total lepton number was an exact
symmetry, while at present, most physicists believe that it is
broken. The process of neutrinoless double beta decay probes the
scale of lepton number symmetry breaking. The standard model does
not conserve $L$ or $B$ separately, but conserves their difference
$B-L$. In a theoretical framework in which quarks and leptons are
unified, one can get $|\Delta B|=2$ processes. This happens, for
example, in a class of unification models where the three colors of
quarks combine with a lepton index to be part of a higher symmetry
group $SU(4)$ which contains the familiar gauge force of strong
interactions based on the color gauge group $SU(3)_c$. Such a
framework automatically leads to the process of $n-\bar{n}$
oscillation without inducing proton decay. The scales of neutrino
mass physics and of $n-\bar{n}$ oscillations are then essentially
the same, and the observation of the latter will provide important
complementary information about the detailed nature of the physics
of neutrino mass. There exist plausible models where this connection
is clearly visible and where the $n-\bar{n}$ oscillation time is
accessible to planned searches. For $n-\bar{n}$ oscillations to be
observable, the scale of $B-L$ breaking must be in the 100 TeV
range, which is fundamentally different from the popular approach
based on grand unified theories (GUT's), where this breaking scale
is around $10^{16}$ GeV.  Thus, an observation of $n-\bar{n}$
oscillations would force us to fundamentally alter our thinking
about unification of forces away from the conventional GUT
approaches to partial unification at intermediate scales or possibly
new physics at the TeV scale, such as TeV$^{-1}$-sized extra
dimensions.

\item {\bf  $n-\bar{n}$ oscillation as a
probe of extra dimensions} \vspace*{0.15in}

There is currently a great deal of interest in the possibility that
there may be extra hidden space-like dimensions in nature.
Motivations for this arise from string theory, specifically in
frameworks where the standard model fields themselves, in addition
to gravity modes, propagate in these extra dimensions.  One of the
appeals of these models is that they can provide an explanation of
the observed generational hierarchy in fermion masses; the
differences in mass between various particles, e.g., up, down, and
strange quarks, would be due to the fact that their chiral
components are located at different sites in these extra dimensions.
In such situations, it has been shown quantitatively that, while
proton decay which involves quarks as well as leptons can be
naturally suppressed,  $n-\bar{n}$ oscillation, which involves only
quarks, is generally unsuppressed, and can be in the observable
range.

\item {\bf Implications for cosmology}
\vspace*{0.15in}

The existence of $n-\bar{n}$ oscillations at an observable rate
would also have profound implications for the origin of
matter-antimatter asymmetry and thus the origin of a net excess of
matter in the Universe. This is due to the fact that observable
$n-\bar{n}$ oscillations would imply that process that violate
baryon number would remain in equilibrium to very low temperatures,
thereby erasing any matter-antimatter asymmetry generated in earlier
epochs, as envisioned in currently popular scenarios such as
leptogenesis. This would then require new ways to understand the
origin of matter. Recently, such new mechanisms have been proposed,
which can be independently tested at colliders.  In these models,
baryon asymmetry is induced after the Universe undergoes the
electroweak phase transition.  Successful baryogenesis, in a class
of models of this type, requires that $n-\bar{n}$ oscillations be
observable, with $\tau_{n-\bar{n}}\sim 10^{10}$ sec.

\item {\bf Dark matter and connection with
neutron--anti-neutron oscillations} \vspace*{0.15in}

There are some attractive models where neutron--anti-neutron
oscillations also provide a probe of the nature of dark matter. A
particularly intriguing suggestion is the possible existence of a
parallel Universe with an identical duplicate of the observed matter
and forces.  Such a scenario could emerge from superstring theory.
In that case, one expects all the particles that we know, protons,
neutrons, etc., to have their mirror partners. Then the possibility
arises that the neutron could oscillate into a mirror neutron. The
same experiment that probes for $n-\bar{n}$ oscillations could also
probe for such processes.  Hydrogen atoms composed of mirror
particles would serve as the dark matter of the Universe in this
scenario.  A possible connection between $n-\bar{n}$ oscillations
and a dark matter candidate arises in models of low-scale
baryogenesis which also have supersymmetry.  The lightest stable
superparticle in these models is not the neutralino, but the partner
of the particle that generated the baryon asymmetry.
\end{enumerate}

\subsection{Experimental Setup at DUSEL}

DUSEL, with its Vertical Facility, can provide a unique opportunity
to advance the search for $n-\bar{n}$ transitions by a sensitivity
factor more than 1000 as compared with the present experimental
limits. This sensitivity will be equivalent to reaching the lifetime
for internuclear $n-\bar{n}$ transition of $\tau_{nuc} \sim 1 \times
10^{35}$ years. The major advantage of the vertical layout as
compared with the alternative approach having a horizontal layout is
the mitigation of the effect of the Earth's gravity on the motion of
cold neutrons over the long flight length.  In the baseline
$n-\bar{n}$ experimental configuration, a 1-km long vertical shaft
of 5-7 meters diameter would be equipped with a vacuum tube and
Earth magnetic field compensation system. A 3.5 MW research reactor
of TRIGA type operating in steady-state mode and installed on the
top of the shaft would serve as the source of neutrons. Neutrons
would be slowed down by a cryogenic liquid deuterium moderator to
typical velocities below 1 km/s and dropped from the top of the
vacuum tube through the focusing supermirror reflector system on an
annihilation detector located at the bottom of the vertical tube.
The background rate in the DUSEL $n-\bar{n}$ detector would be
extremely low, allowing a single observed event to be a discovery.
Active magnetic shielding of the flight tube would allow on/off
switching of the $n-\bar{n}$ transitions if the latter are observed.
The proposed $n-\bar{n}$ experimental configuration is based on
well-established technologies; the main challenge is in the
engineering and vertical construction of the experiment.  Further
factors that can enhance the sensitivity of the vertical $n-\bar{n}$
search are a larger shaft length, larger reflection range of
supermirrors (recently developed at KEK, Japan), development of a
new ``very cold" cryogenic moderator, and higher-power research
reactor.  After three years of running, $n-\overline{n}$ oscillation time sensitivity
will improve by a factor of 1000 compared to the present sensitivity, with
the possibility of improving by an additional factor of 4 in 12 years of
running.

The 3.5 MW TRIGA reactor at the DUSEL Vertical Facility can be
installed on the surface at the distance of about 2 km from the main
underground experimental campus. The antineutrino flux produced by
the reactor can be easily estimated as $\sim$ 62 antineutrinos per
kiloton-year (e.g. by rescaling from the KamLAND detector, where
reactors with 120 GW thermal power at the average distance of 180 km
produce $\sim$ 263 antineutrino events per kiloton-year). This
antineutrino flux certainly can be an essential background for
geo-neutrino detection experiment at DUSEL, but, due to its
controllable nature, it can be precisely measured. The flux of solar
neutrinos to be coped with by the major experiments at the
underground DUSEL site will be substantially larger than the flux of
TRIGA antineutrinos. Given the large distance between the
underground campus and the reactor, the background of thermal
neutrons produced by the TRIGA reactor can be efficiently reduced to
the level of environmental thermal neutron flux by simple passive
shielding.

{\bf Acknowledgements:}  The workshop that generated this paper was
hosted and supported by the Ohio State University Center for
Cosmology and Astro-Particle Physics (CCAPP) and the National
Science Foundation.

\end{document}